\documentclass{ptephy_v1}%%%%%% to generate preprint number with ptep logo

\preprintnumber{XXXX-XXXX} %%% %%% Insert preprint number here
\providecommand{\nubar}{\ensuremath{\bar{\nu}}\xspace}
\providecommand{\nue}{\ensuremath{\nu_{e}}\xspace}
\providecommand{\numu}{\ensuremath{\nu_{\mu}}\xspace}

\providecommand{\nueb}{\ensuremath{\nubar_{e}}\xspace}
\providecommand{\numub}{\ensuremath{\nubar_{\mu}}\xspace}

\providecommand{\nub}{\nubar}

\providecommand{\deltacp}{\ensuremath{\delta_{\scriptscriptstyle\mathrm{CP}}}\xspace}
\providecommand{\ssqthonethree}{\ensuremath{\sin^2\theta_{13}}\xspace}
\providecommand{\ssqthtwothree}{\ensuremath{\sin^2\theta_{23}}\xspace}

\usepackage{hyperref}
\usepackage[T1]{fontenc} % to get the < and > symbols properly
\usepackage{siunitx}
\usepackage{xspace}
\usepackage{multicol}
\usepackage{multirow}
\usepackage{array}
\usepackage{tikz}
\usepackage{booktabs} 
\usepackage[normalem]{ulem} % for strikethrough
\usepackage{xcolor} 
\usepackage{placeins}
\usepackage{multirow} % in preamble
\usepackage{makecell}
\usepackage{subcaption}
\usepackage{graphics} %for dealing with figures.
\title{Updated T2K measurement of neutrino oscillation parameters with the first data from the Gadolinium-loaded far detector}
\author{
\name{\fname{K.} \surname{Abe}}{60}
\name{\fname{S.} \surname{Abe}}{60}
\name{\fname{H.} \surname{Adhikary}}{68}
\name{\fname{R.} \surname{Akutsu}}{15}
\name{\fname{H.} \surname{Alarakia-Charles}}{33}
\name{\fname{Y.I.} \surname{Alj Hakim}}{54}
\name{\fname{S.} \surname{Alonso Monsalve}}{8}
\name{\fname{L.} \surname{Anthony}}{20}
\name{\fname{S.} \surname{Aoki}}{31}
\name{\fname{K.A.} \surname{Apte}}{20}
\name{\fname{T.} \surname{Arai}}{59}
\name{\fname{T.} \surname{Arihara}}{63}
\name{\fname{S.} \surname{Arimoto}}{32}
\name{\fname{Y.} \surname{Asami}}{63}
\name{\fname{Y.} \surname{Asaoka}}{60}
\name{\fname{Y.} \surname{Ashida}}{70}
\name{\fname{E.T.} \surname{Atkin}}{20}
\name{\fname{N.} \surname{Babu}}{37}
\name{\fname{V.} \surname{Baranov}}{38}
\name{\fname{G.J.} \surname{Barker}}{71}
\name{\fname{G.} \surname{Barr}}{46}
\name{\fname{D.} \surname{Barrow}}{46}
\name{\fname{P.} \surname{Bates}}{36}
\name{\fname{L.} \surname{Bathe-Peters}}{46}
\name{\fname{M.} \surname{Batkiewicz-Kwasniak}}{14}
\name{\fname{N.} \surname{Baudis}}{46}
\name{\fname{A.} \surname{Beliakova}}{25}
\name{\fname{V.} \surname{Berardi}}{21}
\name{\fname{L.} \surname{Berns}}{70}
\name{\fname{S.} \surname{Bhattacharjee}}{37}
\name{\fname{A.} \surname{Blanchet}}{57}
\name{\fname{A.} \surname{Blondel}}{57,11}
\name{\fname{L.} \surname{B{\o}e}}{16}
\name{\fname{P.M.M.} \surname{Boistier}}{5}
\name{\fname{S.} \surname{Bolognesi}}{5}
\name{\fname{B.} \surname{Bombin}}{25}
\name{\fname{S.} \surname{Bordoni}}{11}
\name{\fname{S.B.} \surname{Boyd}}{71}
\name{\fname{C.} \surname{Bronner}}{73}
\name{\fname{A.} \surname{Bubak}}{55}
\name{\fname{M.} \surname{Buizza Avanzini}}{35}
\name{\fname{J.A.} \surname{Caballero}}{53}
\name{\fname{N.F.} \surname{Calabria}}{21}
\name{\fname{S.} \surname{Cao}}{19}
\name{\fname{S.} \surname{Cap}}{11}
\name{\fname{D.} \surname{Carabadjac}}{35,5}
\name{\fname{S.L.} \surname{Cartwright}}{54}
\name{\fname{M.P.} \surname{Casado}}{17,75}
\name{\fname{M.G.} \surname{Catanesi}}{21}
\name{\fname{J.} \surname{Chakrani}}{34}
\name{\fname{A.} \surname{Chalumeau}}{57}
\name{\fname{D.} \surname{Cherdack}}{16}
\name{\fname{A.} \surname{Chvirova}}{25}
\name{\fname{J.} \surname{Coleman}}{36}
\name{\fname{G.} \surname{Collazuol}}{23}
\name{\fname{F.} \surname{Cormier}}{65}
\name{\fname{A.A.L.} \surname{Craplet}}{20}
\name{\fname{A.} \surname{Cudd}}{6}
\name{\fname{D.} \surname{D'Ago}}{23}
\name{\fname{C.} \surname{Dalmazzone}}{57}
\name{\fname{T.} \surname{Daret}}{5}
\name{\fname{P.} \surname{Dasgupta}}{17}
\name{\fname{C.} \surname{Davis}}{47}
\name{\fname{Yu.I.} \surname{Davydov}}{38}
\name{\fname{P.} \surname{de Perio}}{28}
\name{\fname{G.} \surname{De Rosa}}{22}
\name{\fname{T.} \surname{Dealtry}}{33}
\name{\fname{C.} \surname{Densham}}{43}
\name{\fname{A.} \surname{Dergacheva}}{25}
\name{\fname{R.} \surname{Dharmapal Banerjee}}{72}
\name{\fname{F.} \surname{Di Lodovico}}{30}
\name{\fname{G.} \surname{Diaz Lopez}}{57}
\name{\fname{S.} \surname{Dolan}}{10}
\name{\fname{D.} \surname{Douqa}}{11}
\name{\fname{T.A.} \surname{Doyle}}{46}
\name{\fname{O.} \surname{Drapier}}{35}
\name{\fname{K.E.} \surname{Duffy}}{46}
\name{\fname{J.} \surname{Dumarchez}}{57}
\name{\fname{P.} \surname{Dunne}}{20}
\name{\fname{K.} \surname{Dygnarowicz}}{69}
\name{\fname{A.} \surname{Eguchi}}{59}
\name{\fname{M.} \surname{El Baz}}{11}
\name{\fname{J.} \surname{Elias}}{49}
\name{\fname{S.} \surname{Emery-Schrenk}}{5}
\name{\fname{G.} \surname{Erofeev}}{25}
\name{\fname{A.} \surname{Ershova}}{35}
\name{\fname{G.} \surname{Eurin}}{5}
\name{\fname{M.} \surname{Fani}}{52}
\name{\fname{D.} \surname{Fedorova}}{25}
\name{\fname{S.} \surname{Fedotov}}{25}
\name{\fname{M.} \surname{Feltre}}{23}
\name{\fname{C.} \surname{Fengler}}{46}
\name{\fname{L.} \surname{Feng}}{32}
\name{\fname{D.} \surname{Ferlewicz}}{57}
\name{\fname{A.J.} \surname{Finch}}{33}
\name{\fname{M.D.} \surname{Fitton}}{43}
\name{\fname{C.} \surname{Forza}}{23}
\name{\fname{M.} \surname{Friend}}{15,76}
\name{\fname{Y.} \surname{Fujii}}{15,76}
\name{\fname{Y.} \surname{Fukuda}}{40}
\name{\fname{N.} \surname{Funayama}}{45}
\name{\fname{Y.} \surname{Furui}}{63}
\name{\fname{A.N.} \surname{Gaci\~no Olmedo}}{57}
\name{\fname{J.} \surname{Garc\'ia-Marcos}}{13}
\name{\fname{A.C.} \surname{Germer}}{47}
\name{\fname{L.} \surname{Giannessi}}{11}
\name{\fname{C.} \surname{Giganti}}{57}
\name{\fname{M.} \surname{Girgus}}{68}
\name{\fname{V.} \surname{Glagolev}}{38}
\name{\fname{P.} \surname{Golak}}{68}
\name{\fname{M.} \surname{Gonin}}{27}
\name{\fname{R.} \surname{Gonzalez Jimenez}}{53}
\name{\fname{J.} \surname{Gonz\'alez Rosa}}{53}
\name{\fname{E.A.G.} \surname{Goodman}}{12}
\name{\fname{K.} \surname{Gorshanov}}{25}
\name{\fname{P.} \surname{Govindaraj}}{68}
\name{\fname{M.} \surname{Grassi}}{23}
\name{\fname{M.} \surname{Guigue}}{57}
\name{\fname{F.Y.} \surname{Guo}}{42}
\name{\fname{D.R.} \surname{Hadley}}{71}
\name{\fname{S.} \surname{Han}}{32,61}
\name{\fname{D.A.} \surname{Harris}}{74}
\name{\fname{R.J.} \surname{Harris}}{33,43}
\name{\fname{M.} \surname{Hartz}}{65,28}
\name{\fname{K.} \surname{Hasegawa}}{32}
\name{\fname{T.} \surname{Hasegawa}}{15,76}
\name{\fname{C.M.} \surname{Hasnip}}{10}
\name{\fname{S.} \surname{Hassani}}{5}
\name{\fname{N.C.} \surname{Hastings}}{15}
\name{\fname{K.} \surname{Hayashi}}{32}
\name{\fname{Y.} \surname{Hayato}}{60,28}
\name{\fname{I.} \surname{Heitkamp}}{70}
\name{\fname{D.} \surname{Henaff}}{5}
\name{\fname{Y.} \surname{Hino}}{15}
\name{\fname{K.} \surname{Hiraide}}{60,28}
\name{\fname{J.} \surname{Holeczek}}{55}
\name{\fname{A.} \surname{Holin}}{43}
\name{\fname{T.} \surname{Holvey}}{46}
\name{\fname{N.T.} \surname{Hong Van}}{26}
\name{\fname{T.} \surname{Honjo}}{45}
\name{\fname{M.C.F.} \surname{Hooft}}{13}
\name{\fname{K.} \surname{Hosokawa}}{60}
\name{\fname{R.} \surname{Huang}}{34}
\name{\fname{J.} \surname{Hu}}{32}
\name{\fname{A.K.} \surname{Ichikawa}}{70}
\name{\fname{K.} \surname{Ieki}}{60}
\name{\fname{M.} \surname{Ikeda}}{60}
\name{\fname{K.} \surname{Indo}}{32}
\name{\fname{T.H.} \surname{Ishida}}{70}
\name{\fname{T.} \surname{Ishida}}{15,76}
\name{\fname{M.} \surname{Ishitsuka}}{64}
\name{\fname{H.} \surname{Ito}}{31}
\name{\fname{S.} \surname{Ito}}{73}
\name{\fname{A.} \surname{Izmaylov}}{25}
\name{\fname{N.} \surname{Jachowicz}}{13}
\name{\fname{B.} \surname{Jargowsky}}{2}
\name{\fname{S.J.} \surname{Jenkins}}{36}
\name{\fname{C.} \surname{Jes\'us-Valls}}{10}
\name{\fname{M.} \surname{Jia}}{42}
\name{\fname{J.J.} \surname{Jiang}}{42}
\name{\fname{J.Y.} \surname{Ji}}{42}
\name{\fname{T.P.} \surname{Jones}}{33}
\name{\fname{P.} \surname{Jonsson}}{20}
\name{\fname{S.} \surname{Joshi}}{42}
\name{\fname{C.K.} \surname{Jung}}{42,28}
\name{\fname{M.} \surname{Kabirnezhad}}{46}
\name{\fname{A.C.} \surname{Kaboth}}{50}
\name{\fname{K.} \surname{Kadota}}{31}
\name{\fname{H.} \surname{Kakuno}}{63}
\name{\fname{A.} \surname{Kamata}}{63}
\name{\fname{J.} \surname{Kameda}}{60}
\name{\fname{S.} \surname{Karpova}}{11}
\name{\fname{V.S.} \surname{Kasturi}}{11}
\name{\fname{Y.} \surname{Kataoka}}{60}
\name{\fname{T.} \surname{Katori}}{30}
\name{\fname{A.} \surname{Kawabata}}{29}
\name{\fname{R.} \surname{Kawabe}}{45}
\name{\fname{Y.} \surname{Kawamura}}{45}
\name{\fname{M.} \surname{Kawaue}}{32}
\name{\fname{E.} \surname{Kearns}}{2,28}
\name{\fname{M.} \surname{Khabibullin}}{25}
\name{\fname{N.V.} \surname{Khomutov}}{38}
\name{\fname{A.} \surname{Khotjantsev}}{25}
\name{\fname{T.} \surname{Kikawa}}{32}
\name{\fname{S.} \surname{King}}{30}
\name{\fname{V.} \surname{Kiseeva}}{38}
\name{\fname{J.} \surname{Kisiel}}{55}
\name{\fname{A.} \surname{Klustov\'a}}{20}
\name{\fname{L.} \surname{Kneale}}{54}
\name{\fname{H.} \surname{Kobayashi}}{59}
\name{\fname{S.R.} \surname{Kobayashi}}{70}
\name{\fname{T.} \surname{Kobayashi}}{15,76}
\name{\fname{L.} \surname{Koch}}{18}
\name{\fname{S.} \surname{Kodama}}{59}
\name{\fname{M.} \surname{Kolupanova}}{25,77}
\name{\fname{A.} \surname{Konaka}}{65}
\name{\fname{L.L.} \surname{Kormos}}{33}
\name{\fname{Y.} \surname{Koshio}}{44,28}
\name{\fname{K.} \surname{Kowalik}}{41}
\name{\fname{R.} \surname{Kralik}}{30}
\name{\fname{Y.} \surname{Kudenko}}{25,77}
\name{\fname{Y.} \surname{Kudo}}{73}
\name{\fname{A.} \surname{Kumar Jha}}{13}
\name{\fname{R.} \surname{Kurjata}}{69}
\name{\fname{V.} \surname{Kurochka}}{25}
\name{\fname{T.} \surname{Kutter}}{37}
\name{\fname{L.} \surname{Labarga}}{1}
\name{\fname{M.} \surname{Lachat}}{49}
\name{\fname{K.} \surname{Lachner}}{8}
\name{\fname{J.} \surname{Lagoda}}{41}
\name{\fname{S.M.} \surname{Lakshmi}}{55}
\name{\fname{M.} \surname{Lamers James}}{71}
\name{\fname{A.} \surname{Langella}}{22}
\name{\fname{D.H.} \surname{Langridge}}{50}
\name{\fname{J.-F.} \surname{Laporte}}{5}
\name{\fname{D.} \surname{Last}}{49}
\name{\fname{N.} \surname{Latham}}{30}
\name{\fname{M.} \surname{Laveder}}{23}
\name{\fname{L.} \surname{Lavitola}}{22}
\name{\fname{M.} \surname{Lawe}}{33}
\name{\fname{A.} \surname{Leclerc}}{5}
\name{\fname{N.} \surname{Lemaire}}{35}
\name{\fname{D.} \surname{Leon Silverio}}{58}
\name{\fname{T.} \surname{Leplumey}}{35}
\name{\fname{S.} \surname{Levorato}}{23}
\name{\fname{S.V.} \surname{Lewis}}{30}
\name{\fname{B.} \surname{Li}}{8}
\name{\fname{C.} \surname{Lin}}{20}
\name{\fname{R.P.} \surname{Litchfield}}{12}
\name{\fname{S.L.} \surname{Liu}}{42}
\name{\fname{W.} \surname{Li}}{46}
\name{\fname{A.} \surname{Longhin}}{23}
\name{\fname{A.} \surname{Lopez Moreno}}{30}
\name{\fname{L.} \surname{Ludovici}}{24}
\name{\fname{X.} \surname{Lu}}{71}
\name{\fname{T.} \surname{Lux}}{17}
\name{\fname{L.N.} \surname{Machado}}{12}
\name{\fname{L.} \surname{Magaletti}}{21}
\name{\fname{K.} \surname{Mahn}}{39}
\name{\fname{K.K.} \surname{Mahtani}}{42}
\name{\fname{S.} \surname{Manly}}{49}
\name{\fname{A.D.} \surname{Marino}}{6}
\name{\fname{D.G.R.} \surname{Martin}}{20}
\name{\fname{D.A.} \surname{Martinez Caicedo}}{58}
\name{\fname{L.} \surname{Martinez}}{17}
\name{\fname{M.} \surname{Martini}}{57,78}
\name{\fname{N.} \surname{Mashin}}{25}
\name{\fname{T.} \surname{Matsubara}}{15}
\name{\fname{R.} \surname{Matsumoto}}{62}
\name{\fname{V.} \surname{Matveev}}{25}
\name{\fname{C.} \surname{Mauger}}{47}
\name{\fname{K.} \surname{Mavrokoridis}}{36}
\name{\fname{N.} \surname{McCauley}}{36}
\name{\fname{K.S.} \surname{McFarland}}{49}
\name{\fname{C.} \surname{McGrew}}{42}
\name{\fname{J.} \surname{McKean}}{32}
\name{\fname{A.} \surname{Mefodiev}}{25}
\name{\fname{G.D.} \surname{Megias}}{53}
\name{\fname{L.} \surname{Mellet}}{39}
\name{\fname{C.} \surname{Metelko}}{36}
\name{\fname{M.} \surname{Mezzetto}}{23}
\name{\fname{S.} \surname{Miki}}{60}
\name{\fname{V.} \surname{Mikola}}{12}
\name{\fname{E.W.} \surname{Miller}}{20}
\name{\fname{A.} \surname{Minamino}}{73}
\name{\fname{O.} \surname{Mineev}}{25}
\name{\fname{S.} \surname{Mine}}{60,4}
\name{\fname{J.} \surname{Mirabito}}{2}
\name{\fname{M.} \surname{Miura}}{60,28}
\name{\fname{S.} \surname{Moriyama}}{60,28}
\name{\fname{S.} \surname{Moriyama}}{73}
\name{\fname{P.} \surname{Morrison}}{12}
\name{\fname{Th.A.} \surname{Mueller}}{35}
\name{\fname{D.} \surname{Munford}}{16}
\name{\fname{A.} \surname{Mu\~noz}}{18}
\name{\fname{L.} \surname{Munteanu}}{10}
\name{\fname{Y.} \surname{Nagai}}{15}
\name{\fname{T.} \surname{Nakadaira}}{15,76}
\name{\fname{K.} \surname{Nakagiri}}{60}
\name{\fname{M.} \surname{Nakahata}}{60,28}
\name{\fname{Y.} \surname{Nakajima}}{59}
\name{\fname{K.D.} \surname{Nakamura}}{70}
\name{\fname{A.} \surname{Nakano}}{70}
\name{\fname{Y.} \surname{Nakano}}{66}
\name{\fname{S.} \surname{Nakayama}}{60,28}
\name{\fname{T.} \surname{Nakaya}}{32,28}
\name{\fname{K.} \surname{Nakayoshi}}{15,76}
\name{\fname{C.E.R.} \surname{Naseby}}{20}
\name{\fname{D.T.} \surname{Nguyen}}{9}
\name{\fname{V.Q.} \surname{Nguyen}}{35}
\name{\fname{K.} \surname{Niewczas}}{35,13}
\name{\fname{S.} \surname{Nishimori}}{59}
\name{\fname{Y.} \surname{Nishimura}}{29}
\name{\fname{Y.} \surname{Noguchi}}{60}
\name{\fname{T.} \surname{Nosek}}{41}
\name{\fname{F.} \surname{Nova}}{43}
\name{\fname{J.C.} \surname{Nugent}}{20}
\name{\fname{H.M.} \surname{O'Keeffe}}{33}
\name{\fname{L.} \surname{O'Sullivan}}{18}
\name{\fname{R.} \surname{Okazaki}}{29}
\name{\fname{W.} \surname{Okinaga}}{59}
\name{\fname{K.} \surname{Okumura}}{61,28}
\name{\fname{T.} \surname{Okusawa}}{45}
\name{\fname{N.} \surname{Onda}}{32}
\name{\fname{N.} \surname{Ospina}}{21}
\name{\fname{L.} \surname{Osu}}{35}
\name{\fname{N.} \surname{Otani}}{32}
\name{\fname{Y.} \surname{Oyama}}{15,76}
\name{\fname{V.} \surname{Paolone}}{48}
\name{\fname{J.} \surname{Pasternak}}{20}
\name{\fname{D.} \surname{Payne}}{36}
\name{\fname{T.P.D.} \surname{Peacock}}{54}
\name{\fname{M.} \surname{Pfaff}}{20}
\name{\fname{L.} \surname{Pickering}}{43}
\name{\fname{J.-B.} \surname{Plan\c{c}on}}{35}
\name{\fname{P.} \surname{Podlaski}}{15,76}
\name{\fname{B.} \surname{Popov}}{57,38}
\name{\fname{A.J.} \surname{Portocarrero Yrey}}{15}
\name{\fname{M.} \surname{Posiadala-Zezula}}{68}
\name{\fname{Y.S.} \surname{Prabhu}}{68}
\name{\fname{H.} \surname{Prasad}}{72}
\name{\fname{F.} \surname{Pupilli}}{23}
\name{\fname{B.} \surname{Quilain}}{27,35}
\name{\fname{P.T.} \surname{Quyen}}{19,79}
\name{\fname{E.} \surname{Radicioni}}{21}
\name{\fname{B.} \surname{Radics}}{74}
\name{\fname{M.A.} \surname{Ramirez Delgado}}{47}
\name{\fname{R.} \surname{Ramsden}}{30}
\name{\fname{P.N.} \surname{Ratoff}}{33}
\name{\fname{M.} \surname{Reh}}{6}
\name{\fname{G.} \surname{Reina}}{18}
\name{\fname{L.} \surname{Restrepo}}{57}
\name{\fname{C.} \surname{Riccio}}{42}
\name{\fname{D.W.} \surname{Riley}}{12}
\name{\fname{E.} \surname{Rondio}}{41}
\name{\fname{D.} \surname{Ross}}{39}
\name{\fname{S.} \surname{Roth}}{51}
\name{\fname{N.} \surname{Roy}}{74}
\name{\fname{A.} \surname{Rubbia}}{8}
\name{\fname{L.} \surname{Russo}}{57}
\name{\fname{A.} \surname{Rychter}}{69}
\name{\fname{W.} \surname{Saenz}}{57}
\name{\fname{K.} \surname{Sakashita}}{67,15,76}
\name{\fname{S.} \surname{Samani}}{11}
\name{\fname{F.} \surname{S\'anchez}}{11}
\name{\fname{E.M.} \surname{Sandford}}{36}
\name{\fname{Y.} \surname{Sato}}{64}
\name{\fname{T.} \surname{Schefke}}{37}
\name{\fname{K.} \surname{Scholberg}}{7,28}
\name{\fname{M.} \surname{Scott}}{20}
\name{\fname{Y.} \surname{Seiya}}{45}
\name{\fname{T.} \surname{Sekiguchi}}{15,76}
\name{\fname{H.} \surname{Sekiya}}{60,28}
\name{\fname{M.} \surname{Sekiyama}}{59}
\name{\fname{T.} \surname{Sekiya}}{63}
\name{\fname{D.} \surname{Seppala}}{39}
\name{\fname{D.} \surname{Sgalaberna}}{8}
\name{\fname{A.} \surname{Shaikhiev}}{25}
\name{\fname{M.} \surname{Shiozawa}}{60,28}
\name{\fname{Y.} \surname{Shiraishi}}{44}
\name{\fname{N.} \surname{Shvarev}}{25}
\name{\fname{A.} \surname{Shvartsman}}{25}
\name{\fname{V.} \surname{Siccardi}}{30}
\name{\fname{N.} \surname{Skrobova}}{25}
\name{\fname{K.} \surname{Skwarczynski}}{41}
\name{\fname{D.} \surname{Smyczek}}{51}
\name{\fname{M.} \surname{Smy}}{4}
\name{\fname{J.T.} \surname{Sobczyk}}{72}
\name{\fname{H.} \surname{Sobel}}{4,28}
\name{\fname{F.J.P.} \surname{Soler}}{12}
\name{\fname{A.J.} \surname{Speers}}{33}
\name{\fname{R.} \surname{Spina}}{21}
\name{\fname{A.} \surname{Srivastava}}{18}
\name{\fname{P.} \surname{Stowell}}{54}
\name{\fname{Y.} \surname{Stroke}}{25}
\name{\fname{I.A.} \surname{Suslov}}{38}
\name{\fname{A.} \surname{Suzuki}}{31}
\name{\fname{M.} \surname{Suzuki}}{73}
\name{\fname{S.Y.} \surname{Suzuki}}{15,76}
\name{\fname{M.} \surname{Tada}}{15,76}
\name{\fname{S.} \surname{Tairafune}}{70}
\name{\fname{A.} \surname{Takeda}}{60}
\name{\fname{Y.} \surname{Takeuchi}}{31,28}
\name{\fname{K.} \surname{Takeya}}{44}
\name{\fname{H.K.} \surname{Tanaka}}{60,28}
\name{\fname{H.} \surname{Tanigawa}}{29}
\name{\fname{A.} \surname{Teklu}}{42}
\name{\fname{V.V.} \surname{Tereshchenko}}{38}
\name{\fname{N.} \surname{Thamm}}{51}
\name{\fname{C.} \surname{Touramanis}}{36}
\name{\fname{N.} \surname{Tran}}{19}
\name{\fname{T.} \surname{Tsukamoto}}{15,76}
\name{\fname{M.} \surname{Tzanov}}{37}
\name{\fname{Y.} \surname{Uchida}}{20}
\name{\fname{M.} \surname{Vagins}}{28,4}
\name{\fname{M.} \surname{Varghese}}{17}
\name{\fname{I.} \surname{Vasilyev}}{38}
\name{\fname{G.} \surname{Vasseur}}{5}
\name{\fname{E.} \surname{Villa}}{8}
\name{\fname{U.} \surname{Virginet}}{57}
\name{\fname{T.} \surname{Vladisavljevic}}{43}
\name{\fname{T.} \surname{Wachala}}{14}
\name{\fname{S.-i.} \surname{Wada}}{31}
\name{\fname{D.} \surname{Wakabayashi}}{70}
\name{\fname{H.T.} \surname{Wallace}}{54}
\name{\fname{J.G.} \surname{Walsh}}{39}
\name{\fname{D.} \surname{Wark}}{43,46}
\name{\fname{M.O.} \surname{Wascko}}{46,43}
\name{\fname{A.} \surname{Weber}}{18}
\name{\fname{R.} \surname{Wendell}}{32}
\name{\fname{M.J.} \surname{Wilking}}{52}
\name{\fname{C.} \surname{Wilkinson}}{34}
\name{\fname{J.R.} \surname{Wilson}}{30}
\name{\fname{C.} \surname{Winterstein}}{5}
\name{\fname{K.} \surname{Wood}}{34}
\name{\fname{C.} \surname{Wret}}{20}
\name{\fname{J.} \surname{Xia}}{56}
\name{\fname{Z.} \surname{Xie}}{52}
\name{\fname{K.} \surname{Yamamoto}}{45}
\name{\fname{T.} \surname{Yamamoto}}{45}
\name{\fname{T.} \surname{Yamazumi}}{59}
\name{\fname{C.} \surname{Yanagisawa}}{42,80}
\name{\fname{Y.} \surname{Yang}}{46}
\name{\fname{T.} \surname{Yano}}{60}
\name{\fname{N.} \surname{Yershov}}{25}
\name{\fname{U.} \surname{Yevarouskaya}}{42}
\name{\fname{M.} \surname{Yokoyama}}{59,28}
\name{\fname{Y.} \surname{Yoshimoto}}{59}
\name{\fname{N.} \surname{Yoshimura}}{32}
\name{\fname{R.} \surname{Zaki}}{74}
\name{\fname{A.} \surname{Zalewska}}{14}
\name{\fname{J.} \surname{Zalipska}}{41}
\name{\fname{G.} \surname{Zarnecki}}{14}
\name{\fname{J.} \surname{Zhang}}{65,3}
\name{\fname{X.Y.} \surname{Zhao}}{20}
\name{\fname{H.} \surname{Zheng}}{42}
\name{\fname{H.} \surname{Zhong}}{31}
\name{\fname{T.} \surname{Zhu}}{20}
\name{\fname{M.} \surname{Ziembicki}}{69}
\name{\fname{E.D.} \surname{Zimmerman}}{6}
\name{\fname{M.} \surname{Zito}}{57}
\name{\fname{S.} \surname{Zsoldos}}{30}
}

\address{
\affil{1}{University Autonoma Madrid, Department of Theoretical Physics, 28049 Madrid, Spain}
\affil{2}{Boston University, Department of Physics, Boston, Massachusetts, U.S.A.}
\affil{3}{University of British Columbia, Department of Physics and Astronomy, Vancouver, British Columbia, Canada}
\affil{4}{University of California, Irvine, Department of Physics and Astronomy, Irvine, California, U.S.A.}
\affil{5}{IRFU, CEA, Universit\'e Paris-Saclay, F-91191 Gif-sur-Yvette, France}
\affil{6}{University of Colorado at Boulder, Department of Physics, Boulder, Colorado, U.S.A.}
\affil{7}{Duke University, Department of Physics, Durham, North Carolina, U.S.A.}
\affil{8}{ETH Zurich, Institute for Particle Physics and Astrophysics, Zurich, Switzerland}
\affil{9}{VNU University of Science, Vietnam National University, Hanoi, Vietnam}
\affil{10}{CERN European Organization for Nuclear Research, CH-1211 Gen\'eve 23, Switzerland}
\affil{11}{University of Geneva, Section de Physique, DPNC, Geneva, Switzerland}
\affil{12}{University of Glasgow, School of Physics and Astronomy, Glasgow, United Kingdom}
\affil{13}{Ghent University, Department of Physics and Astronomy, Proeftuinstraat 86, B-9000 Gent, Belgium}
\affil{14}{H. Niewodniczanski Institute of Nuclear Physics PAN, Cracow, Poland}
\affil{15}{High Energy Accelerator Research Organization (KEK), Tsukuba, Ibaraki, Japan}
\affil{16}{University of Houston, Department of Physics, Houston, Texas, U.S.A.}
\affil{17}{Institut de Fisica d'Altes Energies (IFAE) - The Barcelona Institute of Science and Technology, Campus UAB, Bellaterra (Barcelona) Spain}
\affil{18}{Institut f\"ur Physik, Johannes Gutenberg-Universit\"at Mainz, Staudingerweg 7, 55128 Mainz, Germany}
\affil{19}{Institute For Interdisciplinary Research in Science and Education (IFIRSE), ICISE, Quy Nhon, Vietnam}
\affil{20}{Imperial College London, Department of Physics, London, United Kingdom}
\affil{21}{INFN Sezione di Bari and Universit\`a e Politecnico di Bari, Dipartimento Interuniversitario di Fisica, Bari, Italy}
\affil{22}{INFN Sezione di Napoli and Universit\`a di Napoli, Dipartimento di Fisica, Napoli, Italy}
\affil{23}{INFN Sezione di Padova and Universit\`a di Padova, Dipartimento di Fisica, Padova, Italy}
\affil{24}{INFN Sezione di Roma and Universit\`a di Roma ``La Sapienza'', Roma, Italy}
\affil{25}{Institute for Nuclear Research of the Russian Academy of Sciences, Moscow, Russia}
\affil{26}{International Centre of Physics, Institute of Physics (IOP), Vietnam Academy of Science and Technology (VAST), 10 Dao Tan, Ba Dinh, Hanoi, Vietnam}
\affil{27}{ILANCE, CNRS – University of Tokyo International Research Laboratory, Kashiwa, Chiba 277-8582, Japan}
\affil{28}{Kavli Institute for the Physics and Mathematics of the Universe (WPI), The University of Tokyo Institutes for Advanced Study, University of Tokyo, Kashiwa, Chiba, Japan}
\affil{29}{Keio University, Department of Physics, Kanagawa, Japan}
\affil{30}{King's College London, Department of Physics, Strand, London WC2R 2LS, United Kingdom}
\affil{31}{Kobe University, Kobe, Japan}
\affil{32}{Kyoto University, Department of Physics, Kyoto, Japan}
\affil{33}{Lancaster University, School of Physics and Astronomy, Lancaster, United Kingdom}
\affil{34}{Lawrence Berkeley National Laboratory, Berkeley, California, U.S.A.}
\affil{35}{Ecole Polytechnique, IN2P3-CNRS, Laboratoire Leprince-Ringuet, Palaiseau, France}
\affil{36}{University of Liverpool, Department of Physics, Liverpool, United Kingdom}
\affil{37}{Louisiana State University, Department of Physics and Astronomy, Baton Rouge, Louisiana, U.S.A.}
\affil{38}{Joint Institute for Nuclear Research, Dubna, Moscow Region, Russia}
\affil{39}{Michigan State University, Department of Physics and Astronomy,  East Lansing, Michigan, U.S.A.}
\affil{40}{Miyagi University of Education, Department of Physics, Sendai, Japan}
\affil{41}{National Centre for Nuclear Research, Warsaw, Poland}
\affil{42}{State University of New York at Stony Brook, Department of Physics and Astronomy, Stony Brook, New York, U.S.A.}
\affil{43}{STFC, Rutherford Appleton Laboratory, Harwell Oxford,  and  Daresbury Laboratory, Warrington, United Kingdom}
\affil{44}{Okayama University, Department of Physics, Okayama, Japan}
\affil{45}{Osaka Metropolitan University, Department of Physics, Osaka, Japan}
\affil{46}{Oxford University, Department of Physics, Oxford, United Kingdom}
\affil{47}{University of Pennsylvania, Department of Physics and Astronomy,  Philadelphia, Pennsylvania, U.S.A.}
\affil{48}{University of Pittsburgh, Department of Physics and Astronomy, Pittsburgh, Pennsylvania, U.S.A.}
\affil{49}{University of Rochester, Department of Physics and Astronomy, Rochester, New York, U.S.A.}
\affil{50}{Royal Holloway University of London, Department of Physics, Egham, Surrey, United Kingdom}
\affil{51}{RWTH Aachen University, III. Physikalisches Institut, Aachen, Germany}
\affil{52}{School of Physics and Astronomy, University of Minnesota, Minneapolis, Minnesota, U.S.A.}
\affil{53}{Departamento de F\'isica At\'omica, Molecular y Nuclear, Universidad de Sevilla, 41080 Sevilla, Spain}
\affil{54}{University of Sheffield, School of Mathematical and Physical Sciences, Sheffield, United Kingdom}
\affil{55}{University of Silesia, Institute of Physics, Katowice, Poland}
\affil{56}{SLAC National Accelerator Laboratory, Stanford University, Menlo Park, California, U.S.A.}
\affil{57}{Sorbonne Universit\'e, CNRS/IN2P3, Laboratoire de Physique Nucl\'eaire et de Hautes Energies (LPNHE), Paris, France}
\affil{58}{South Dakota School of Mines and Technology, 501 East Saint Joseph Street, Rapid City, SD 57701, United States}
\affil{59}{University of Tokyo, Department of Physics, Tokyo, Japan}
\affil{60}{University of Tokyo, Institute for Cosmic Ray Research, Kamioka Observatory, Kamioka, Japan}
\affil{61}{University of Tokyo, Institute for Cosmic Ray Research, Research Center for Cosmic Neutrinos, Kashiwa, Japan}
\affil{62}{Institute of Science Tokyo, Department of Physics, Tokyo}
\affil{63}{Tokyo Metropolitan University, Department of Physics, Tokyo, Japan}
\affil{64}{Tokyo University of Science, Faculty of Science and Technology, Department of Physics, Noda, Chiba, Japan}
\affil{65}{TRIUMF, Vancouver, British Columbia, Canada}
\affil{66}{University of Toyama, Faculty of Science, Toyama, Japan}
\affil{67}{University of Tsukuba, Faculty of Pure and Applied Sciences, Tsukuba, Ibaraki, Japan}
\affil{68}{University of Warsaw, Faculty of Physics, Warsaw, Poland}
\affil{69}{Warsaw University of Technology, Institute of Radioelectronics and Multimedia Technology, Warsaw, Poland}
\affil{70}{Tohoku University, Faculty of Science, Department of Physics, Miyagi, Japan}
\affil{71}{University of Warwick, Department of Physics, Coventry, United Kingdom}
\affil{72}{Wroclaw University, Faculty of Physics and Astronomy, Wroclaw, Poland}
\affil{73}{Yokohama National University, Department of Physics, Yokohama, Japan}
\affil{74}{York University, Department of Physics and Astronomy, Toronto, Ontario, Canada}
\affil{75}{Departament de Fisica de la Universitat Autonoma de Barcelona, Barcelona, Spain}
\affil{76}{J-PARC, Tokai, Japan}
\affil{77}{Moscow Institute of Physics and Technology (MIPT), Moscow region, Russia and National Research Nuclear University ``MEPhI'', Moscow, Russia}
\affil{78}{IPSA-DRII, France}
\affil{79}{the Graduate University of Science and Technology, Vietnam Academy of Science and Technology}
\affil{80}{BMCC/CUNY, Science Department, New York, New York, U.S.A.}
}

\begin{document}

%%%% To generate auto affiliation numbers please use \author{}\affil{} command
%\author{}
%\author{The T2K Collaboration}\noaffiliation

\date{}

%%% To include the collaborator name... Please use the command "\collaborator"
%%% For example: \collaborator{ATLAS Collaboration}

\begin{abstract}
The T2K experiment presents updated measurements of neutrino oscillation parameters using $21.4(16.3)\times10^{20}$ protons on target (POT) in (anti-)neutrino mode at the far detector, Super-Kamiokande. This analysis uses an additional neutrino dataset of $1.7 \times 10^{20}$ POT compared to the previous result, and for the very first time, data collected with Super-Kamiokande doped with Gadolinium (Gd). %Moreover,
Significant improvements were introduced in the far detector selection and systematic error treatment. %, and
%added to the SK detector selection and systematic error treatment, and %, updating the selection for decay-electrons and the SK detector systematic uncertainty modeling. 
%the robustness of our data fit was additionally verified using a supplementary simulated data set. 
Both semi-frequentist and Bayesian analyses are presented, including results on \ssqthtwothree, $\Delta m^2_{32} $, \ssqthonethree and \deltacp. Both analyses favor normal mass ordering with a nearly maximal CP-violating phase and are compatible with neutrino maximal mixing, slightly favouring the upper $\theta_{23}$  octant. 

%Assuming the normal neutrino mass ordering and using the constraint on \ssqthonethree from reactor experiments, this measurement campaign reports the following central results and $1\sigma$ confidence levels: $\sin^{2} \theta_{23} = 0.568^{+0.014}_{-0.036}$ (note that our data are also compatible with the lower octant region $0.464-0.482$ at $1\sigma$) using Feldman–Cousins corrected intervals, and $\Delta m^{2}_{32} = 2.521^{+0.037}_{-0.050} \times 10^{−3}$~eV$^{2}$ using constant $\Delta \chi^{2}$ intervals. The CP-violating phase is constrained to $\delta_{CP} = −2.08^{+1.33}_{-0.61}$ using Feldman–Cousins corrected intervals. Finally, $\delta_{CP} = 0, \pi$ are excluded at more than $90\%$ confidence level. Similarly to last analysis, a Jarlskog invariant value of zero is excluded at more than $2\sigma$ credible level assuming a flat prior in $\delta_{CP}$, and just below $2\sigma$ with a flat prior in $\sin \delta_{CP}$. Finally, when the external reactor constraint is lifted, our data are compatible with %$\sin^{2}_{\theta_{13}} = 28.0+2.8
%−6.5 × 10−3, in agreement with measurements from reactor experiments. These results are consistent with previous T2K analyses.

\end{abstract}

\subjectindex{xxxx, xxx}
\maketitle

\section{Introduction}
After the very first observation of neutrino oscillations by Super-Kamiokande (SK)~\cite{Super-Kamiokande:1998kpq}, experiments now aim to fully characterize the three-flavor mixing paradigm described by the Pontecorvo–Maki–Nakagawa–Sakata (PMNS) matrix \cite{Pontecorvo:1967fh,MNS10.1143/PTP.28.870}. In this framework, the neutrino mixing is governed by three mixing angles ($\theta_{13}$, $\theta_{23}$, and $\theta_{12}$), two mass square differences ($\Delta {m}^{2}_{32}$ and $\Delta {m}^{2}_{21}$), and at least one Charge Parity (CP) violating phase (\deltacp). Over the past two decades, most of these oscillation parameters have been precisely measured \cite{ParticleDataGroup:2024cfk}. However, the CP-violating phase and the ordering of the neutrino mass states (MO), corresponding to the sign of $\Delta {m}^{2}_{32}$, still remain unresolved. Moreover, improved measurements of the other parameters is essential for precisely testing the PMNS unitarity. Since both the CP-violating phase and the mass ordering affect neutrino ($\nu$) and antineutrino ($\nub$) oscillation probabilities differently, accelerator neutrino experiments capable of producing well-controlled and highly pure $\numu$ and $\numub$ beams provide an ideal approach to probe these effects.

This paper presents an updated measurement of the neutrino oscillation parameters from the long-baseline T2K (Tokai to Kamioka) experiment~\cite{T2K:2011qtm}. It uses $21.4(16.3)\times10^{20}$ protons on target (POT) in (anti-)neutrino mode collected using the Super-Kamiokande far detector (FD).  The new data correspond to an increase of approximately 9\% in the neutrino-mode statistics compared to the previously published paper~\cite{OA22shortPaper},  and were collected for the first time with a 0.01\% gadolinium (Gd) loading in Super-Kamiokande. The enhanced neutron capture efficiency provided by Gd required updated selection criteria to suppress possible backgrounds, as described in Section~\ref{sec:fd}. In addition, a new FD response model has been developed, together with an improved treatment of the covariance matrix encoding event rate uncertainties of each FD oscillation sample and more refined handling of systematic errors, as discussed in Section~\ref{sec:fd_sys}. 

In this paper, the T2K experimental apparatus and data are described in Sec.~\ref{sec:T2K}, including updates to SK for Gd loading. The oscillation analysis method is introduced in Sec~\ref{sec:Method}, along with significant recent improvements to FD data analysis. Finally, the updated oscillation analysis results are presented and discussed in Sec.~\ref{sec:Results}.

 %Neutrino oscillations are...
%The results presented in this paper represent an update to ...
%This paper is organized as follows: ...

\section{The T2K Experiment}
\label{sec:T2K}
%The accurate determination of oscillation parameters relies on the precise control of the produced and detected neutrino flavors, the tuning and knowledge of the neutrino energy and travel distance to maximize the oscillation probability, and a high efficiency in neutrino generation and data collection. 
Accurate determination of neutrino oscillation parameters imposes several stringent requirements on the experiment. The neutrino flavor composition must be precisely known at both production and detection, while the neutrino energy and baseline must be carefully optimized to maximize oscillation sensitivity. In addition, high efficiency in both neutrino production and data acquisition is essential to ensure sufficient statistical precision.
%\sout{For these purposes, T2K relies on an accelerator located in J-PARC that produces a high-purity beam of \numu or \numub with a well-controlled peak energy of $E_{\nu} \sim 0.6$~GeV. This generated neutrino flux is then detected 295~km away at the FD, allowing to measure the $\nue$ (resp. $\nueb$) appearance and \numu (resp. \numub) disappearance from the \numu (resp. \numub) beam. }\textcolor{green}{ all this is repeated in the following, apart the 295 km baseline (Mauro)}
For these purposes, T2K relies on an accelerator located in J-PARC that produces an ultra-pure beam of \numu or \numub. The neutrino beam is generated by a sequence of three accelerator stages to accelerate protons to an energy of \SI{30}{\GeV}, which are then extracted to impinge on a graphite target. Interactions in the target produce hadrons, mostly pions, which are focused using three sequential magnetic horns~\cite{T2K:2015sbp}. The polarity of the magnetic field produced by the horns is switchable, allowing us to select positively (negatively) charged hadrons that decay, creating a \numu (\numub) beam.  The T2K beam is aimed with a 2.5$^{\circ}$ angle off-axis from the direction to SK, at a distance of 295 km. This configuration produces a narrow-band neutrino spectrum peaked at $\sim 0.6$ GeV, corresponding to the first  $\numu \rightarrow \nue$ oscillation maximum, while reducing backgrounds from higher energy neutrino interactions and the contamination from the intrinsic electron neutrino component of the beam.
A suite of near detectors (NDs), located approximately 280~m from the beam production target, allows the characterization of the unoscillated T2K beam. The on-axis near detector INGRID monitors beam intensity and direction \cite{Abe:2011xv,Otani:2010zza}, while the off-axis ND280 detector -- positioned  so as to have the same off-axis angle as SK \textit{i.e.} at \SI{2.5}{\degree} off-axis --measures neutrino interaction rates to constrain systematic uncertainties from flux and cross-sections.
 ND280 is housed within a 0.2 T magnet, previously used by the UA1 and NOMAD experiments at CERN, and comprises two fine-grained carbon-based scintillator detectors that serve as the primary neutrino targets (FGD1 and FGD2), with FGD2 also incorporating water layers \cite{T2KND280FGD:2012umz}. The FGDs are interleaved with three time-projection chambers (TPCs) \cite{T2KND280TPC:2010nnd} and are surrounded by an electromagnetic calorimeter \cite{T2KUK:2013wkh}. An upstream $\pi^0$ detector \cite{Assylbekov:2011sh}, together with side muon range detectors installed on the magnet yoke \cite{Aoki:2013swe}, completes the ND280 detector.

The \nue(\nueb) appearance and \numu(\numub) disappearance are measured at Super-Kamiokande~\cite{Super-Kamiokande:2002weg,Abe:2013gga}, a 50 kt water Cherenkov detector instrumented with 11,129 inward-facing 20-inch photomultiplier tubes (PMTs) that detect Cherenkov radiation emitted by charged particles traversing the detector. Electron and muon neutrinos are identified through the flavor of the charged lepton produced in charged-current interactions in the detector. Appearing as ring-like patterns on the cylindrical inner surface of the detector, the charged-lepton Cherenkov light is identified by its ring profile: electrons produce diffuse rings due to scattering, while muons produce rings with well-defined edges. Using this sole PID method, SK can maintain a $99.9\%$ \nue selection efficiency while suppressing more than $97\%$ of the \numu background.
%~\cite{Super-Kamiokande:2019gzr}.
Michel electrons from muon decays are also used to improve muon event identification. 
%To reject particles generated by interactions outside the SK inner detector (ID), SK is equipped with an outer detector (OD) around the ID, instrumented with 1,885 outward-facing 8-inch PMTs. 
%, mainly by identifying the electron- and muon-like Cherenkov rings, respectively.
%SK is a 50 kt water Cherenkov detector with 11,129 inward-facing 20-inch photomultiplier tubes (PMTs) to detect Cherenkov radiation from charged particles traversing the detector. To reject particles generated by interactions outside the SK inner detector (ID), SK is equipped with an outer detector (OD) around the ID, instrumented with 1,885 outward-facing 8-inch PMTs. 
 %Since 2020, SK has begun dissolving gadolinium sulfate octahydrate (Gd$_2$(SO$_4$)$_3\,\cdot$8H$_2$O) in the detector water to reach a neutron tagging efficiency of 75\% at a Gd concentration of 0.03\%~\cite{Abe_2022_skgd, ABE2024169480_skgd}. 

 In 2020, gadolinium sulfate octahydrate (Gd$_2$(SO$_4$)$_3\,\cdot$8H$_2$O) was dissolved in the SK detector water at a Gd concentration of 0.01\%~\cite{Abe_2022_skgd}. In 2022, the concentration was increased to 0.03\%~\cite{ABE2024169480_skgd}. The addition of gadolinium enhances the detection efficiency of neutron capture to approximately 50\% (and up to 75\% at 0.03\% concentration), thereby improving the detection efficiency for low-energy antineutrinos ~\cite{ABE2024169480_skgd}. 
This modification has a negligible impact on oscillation analyses. However,
   gamma rays from neutron capture on Gd can be misidentified as Michel electrons, potentially affecting the event selection used in T2K. Updated selection criteria have therefore been introduced to mitigate Gd-induced background. More details are provided in Sec.~\ref{sec:fd}. The newly acquired data presented in this paper were taken with a concentration of 0.01\% Gd in SK.

\section{Analysis Method}
\label{sec:Method}
We use two fitters in the oscillation analysis, one Bayesian~\cite{the_mach3_collaboration_2026_20187111} and one semi-frequentist, which will be further described in section \ref{sec:AnaMethod}. The two fitters are extensively cross-validated against each other. In the Bayesian fit, the systematic parameters are constrained simultaneously using data from the near and far detectors. In the semi-frequentist fit, constraints from the near detectors are propagated to the far detector through a covariance matrix. The fitter inputs from the neutrino flux, the interaction models and the near and far detector models are described in the following sections.
%The parameters constrained by the ND fit that are related to the FD are propagated further to the FD fit.

\subsection{Beam flux}
\label{sec:BeamFlux}

The neutrino flux estimation is based on a simulation tuned to hadron production measurements from the NA61/SHINE experiment at CERN \cite{NA61:2014lfx}, which used a 30~GeV proton beam as T2K, hitting thin targets~\cite{Abgrall:2015hmv} and a replica of the T2K target~\cite{Abgrall:2016jif}.  
The details of this tuning are described in~\cite{T2K:2023smv}. The flux model used is the same as in the previous analysis~\cite{OA22shortPaper}, but has been retuned with the additional data corresponding to the entire neutrino data set of this paper \textit{i.e.} $21.4(16.3)\times10^{20}$ POT.
%, and summarized below.
%The incoming protons are generated within the simulation using the proton beam profiles measured for each run.
%Their interactions inside the graphite target (91.6 cm long, 2.6 cm diameter) are simulated with FLUKA version 2011.2x~\cite{Ahdida:2022gjl,BATTISTONI201510}. The particle paths through the three magnetic horns and the 96 m long decay tunnel until their decay into neutrinos are simulated with the GEANT3-based \textsc{Jnubeam} package~\cite{Abe:2012av}. 
%Charged pions exiting from the target are tuned using the replica target measurements. Outgoing particles not fully covered by the replica target measurements, such as kaons, protons and any hadronic interactions outside of the target, are simulated with a tuning derived from thin target measurements.
%The uncertainties on hadronic interactions in T2K are derived from the the statistical and systematic uncertainties on measured yields by NA61/SHINE experiment, while
%the uncertainties of interactions unconstrained by external data are estimated with comparisons between Monte Carlo (MC) hadron interaction models. A covariance matrix of the flux at the near and far detectors for each neutrino flavor in the two beam modes is then used to propagate the neutrino flux constraint at the near detector to the far detector prediction.
 The relative uncertainty of this tuning of the \numu flux in $\nu$-mode and \numub flux in \nubar-mode is estimated to be about 5--8\% near the flux spectrum peak. For the \numub component in $\nu$-mode and \numu component in \nubar-mode (``wrong-sign background''), the uncertainty has a~larger contribution from interactions occurring outside the main target, resulting in an uncertainty of about 6--8\%. Note the contribution of wrong-sign background is larger in \nubar-mode due to the higher \numu charged-current cross-section with matter than \numub.

\subsection{Neutrino interactions}
\label{sec:nuint}
The neutrino interaction and the relevant systematic uncertainty model in this work remain the same as in the last analysis~\cite{OA22shortPaper}. The underlying model relies on the NEUT neutrino-nucleus interaction event generator~\cite{Hayato:2009zz} version 5.4.0. 

The dominant channels at T2K energies are charged-current quasi-elastic (CCQE) and the CC two-particle two-hole (2p2h) described by the ``Benhar Spectral Function'' model~\cite{BENHAR1994493,NEUT-Benhar} and the Nieves et al. model~\cite{Nieves2p2h}, respectively, with associated uncertainties, including parameters to account for the impact of Pauli blocking and nuclear transparency. A complete description of the nuclear model is given in~\cite{NDNIWGpaper}.

Single-pion production (SPP) in this analysis is described by the Rein-Sehgal model~\cite{RS-model}, including resonance interference and coherent scattering based on the Berger-Sehgal model~\cite{BergerSehgal}.

%, with uncertainties derived from \textsc{NEUT}~\cite{Hayato:2009zz} and NuWro~\cite{NuWro} comparison.

Deep inelastic scattering employs GRV98 Parton Distribution Functions (PDFs)~\cite{grv98} with Bodek-Yang (BY) modifications~\cite{BY1,BY2} to extend to low $Q^2$. Effective uncertainties on the BY correction are applied. The uncertainty on the pion multiplicity and invariant hadronic mass distribution is developed by comparing NEUT~\cite{Hayato:2009zz} and the AGKY model~\cite{AGKY}.

\subsection{Near detector}
The near detector (ND) analysis aims to reduce uncertainties in the predicted number of events at the far detector by constraining the unoscillated neutrino flux and interaction cross-section models. This is achieved by fitting ND data binned in muon momentum and in the angle of the outgoing muon with respect to the incoming neutrino.

The ND fit utilizes the same selected event samples and methodology as the previous analysis~\cite{OA22shortPaper}, with the sole modification being a re-tuned beam prediction that incorporates the overall flux uncertainty from the new FD data collected using loaded gadolinium. 

At the ND, CC-inclusive events were selected by identifying the muon candidate as the highest-momentum track originating in either of the two fine-grained detectors (FGD1/FGD2). The selected CC events were then separated into samples according to the FGD in which the interaction occurred, the beam mode, the muon charge and the reconstructed final-state particle multiplicities. Consistent with the previous analysis~\cite{OA22shortPaper}, a total of 22 selected samples were used in the fit.

%In ND, CC-inclusive events were selected by identifying muon candidates as tracks with the highest momentum that start in one of the two fine-grained detectors (FGDs). Furthermore, the selected CC events were separated into different samples according to the FGD in which the interaction occurred, the beam mode, the muon charge and the reconstructed final-state particle multiplicities. Consistent with the previous analysis, a total of 22 selected samples were used in the fit.

%The ND elected CC events were separated into different samples according to the FGD in which the interaction occurred, the beam mode, the muon charge and the reconstructed final-state particle multiplicities. Negatively-charged muon candidates selected in $\nu$-mode were divided into five samples per FGD: 1) at least one photon (CC-photon), 2) no photons, pions or protons (CC0$\pi$0p), 3) no photons or pions and at least one proton (CC0$\pi$Np), 4) no photons and one positively-charged pion (CC1$\pi$+), 5) and all other CC events (CC-Other).
%Positively and negatively charged muon candidates in  ̄ν-mode are each divided into three samples per FGD: 1) no pions (CC0$\pi$), 2) one pion with opposite charge to the muon (CC1$\pi$), 3) and all other CC events (CC-Other). 

%Fig.~\ref{fig:ndsample} illustrates an example of the data and fit result specifically for the $\nu_{\mu}$ CC0$\pi$0p sample.

The ND fitting strategy involves finding the global maximum of likelihood by using the MIGRAD gradient descent method, as implemented in Minuit2~\cite{Hatlo:2005cj},  to ensure that ND event rate predictions match the data. The Hessian matrix estimated through MIGRAD is used to estimate the parameter uncertainties.

%\begin{figure}[htb!]
%    \centering
    %\includegraphics[width=0.49\linewidth]{ND280_figs/pmu_1d_modes_postfit_fgd1_numucc_0pi_0_protons_no_photon.pdf}\includegraphics[width=0.49\linewidth]{ND280_figs/costheta_1d_modes_postfit_fgd1_numucc_0pi_0_protons_no_photon.pdf}
    %\caption{ND data and postfit MC prediction, and their ratio for the neutrino-mode $\nu_{\mu}$ interacting through charged-current with no charged $\pi$ and no proton in the final state (CC0$\pi0$p) with respect to the reconstructed muon momentum (left) and angle with respect to the beam direction (right). The MC prediction is broken down by interaction mode.}
    %\label{fig:ndsample}
%\end{figure}

In the ND analysis, the Barlow-Beeston likelihood with Conway implementation~\cite{Barlow:1993dm, Conway} was used, where prior constraints on flux prediction, detector effects, and neutrino interaction cross-section models enter a~likelihood penalty term and ND neutrino events, split into samples,  form the statistical part.

%In ND, CC-inclusive events are selected by identifying muon candidates as tracks with the highest momentum that start in one of the two fine-grained detectors (FGDs).

%Additionally, based on the particles reconstructed in the final state, the CC-inclusive events were further subdivided into pion-, proton-, and photon-tagged samples. For anti-neutrino events, only pion-tagging was applied. The samples were also categorized according to the fine-grained detector module in which the event interaction vertex was reconstructed, either one of the FGDs. Consistent with the previous analysis, a total of 22 selected samples were used in the fit.

%Parameters 
%\textcolor{green}{ (it's too vague, one could believe that we are talking about oscillation parameters. A little bit of explanation about detector and xsec parameters is necessary, how nominal values are chosen and how many parameters are constrained by the fit (Mauro))} enter the fit at their nominal values, along with their prior uncertainties, correlations, and variation limits.

The ND280 model is described by 727 parameters in total, comprising 100 flux, 552 detector, and 75 cross-section parameters. Flux and detector parameters are treated as normalization parameters, whereas cross-section parameters include both normalization and shape effects. To assess the robustness of the minimization in this high-dimensional parameter space, we performed extensive comparison against the independent Bayesian Markov chain Monte Carlo fitting framework, and found consistent and reasonable agreement in the fitted results.

%Additionally, based on the particles reconstructed in the final state, the CC-inclusive events are further divided into pion-, proton-, and photon-tagged samples. In anti-neutrino samples, only pion-tagging is applied. The samples are further categorized into FGD1 and FGD2-related events. As in the previous analysis, in total 22 selected samples are used in the fit.

%Parameters enter the fit at their nominal values, along with their prior uncertainties, correlations and variations limits.

%Since no new ND data entering the fit and the only update is updated overall flux covariance was used in the fit that includes data collected using loaded gadolinium, the impact of the new flux constraint on near detector predictions was negligible compared to the previous analysis~\cite{OA22shortPaper}.

Since no new ND data has entered the fit compared with the previous analysis~\cite{OA22shortPaper}, and the only update is the inclusion of a new flux constraint, the impact of this constraint on near-detector predictions is, as expected, negligible.

%The results of the ND fit show that the inclusion of gadolinium data has a minimal effect on the ND constraints calculations compared to the previous analysis. Consequently, the p-value remains the same as in the previous analysis. 

\subsection{Far detector}
\label{sec:fd}
At the FD, neutrino interactions are reconstructed by analysing the pattern of Cherenkov light recorded by the detector's PMTs. A maximum-likelihood framework 
%\cite{10.1093/ptep/ptz015}
\cite{Super-Kamiokande:2019gzr}
is used to determine the number of light-emitting particle tracks (rings), identify particle types (PID), and estimate their kinematics.

Six signal-enriched samples are used in this analysis, which are the same as in the previous one \cite{OA22shortPaper}. Five of these are identified by a single ring-like pattern from the prompt PMT hits. These are CCQE-enhanced $\mu$-like and $e$-like single-ring (1R) events collected with the beam running in the $\nu$ and $\bar\nu$ modes, and the CC single-pion production dominant 1R$e$-like sample with one decay electron (1Re1d.e.), collected only with the $\nu$ beam. 
%The sixth sample targets $\nu_\mu$ CC1$\pi^+$ interactions by selecting both single and multi-ring $\nu_\mu$ events with at least one decay electron. 
The sixth sample targets $\nu_\mu$ CC1$\pi^+$ interactions, identified by multi-ring $\mu$-like events with at least one decay electron or single-ring $\mu$-like events with at least two decay electrons. 
 
Table \ref{tab:run1-11_data} summarizes the number of neutrino events across SK selections in our data and model for different $\delta_{CP}$ values.

Secondary particles, such as Michel electrons, generate delayed PMT hit clusters. In T2K's previous analyses, Michel electrons at SK were identified by counting these time-separated PMT hit clusters.

Neutron captures also produce delayed hits, typically occurring well beyond the muon decay time of 2.2~$\mu$s (at characteristic times of 200~$\mu$s in pure water and 30~$\mu$s with Gd loading). Because neutron captures produce much less light than Michel electrons, the previous identification algorithm experienced a low contamination—about $5\%$—from neutron captures on water among Michel electron candidates.

\begin{figure}[htb!]
    \centering

    \begin{subfigure}{0.32\linewidth}
        \centering
        \includegraphics[width=\linewidth]{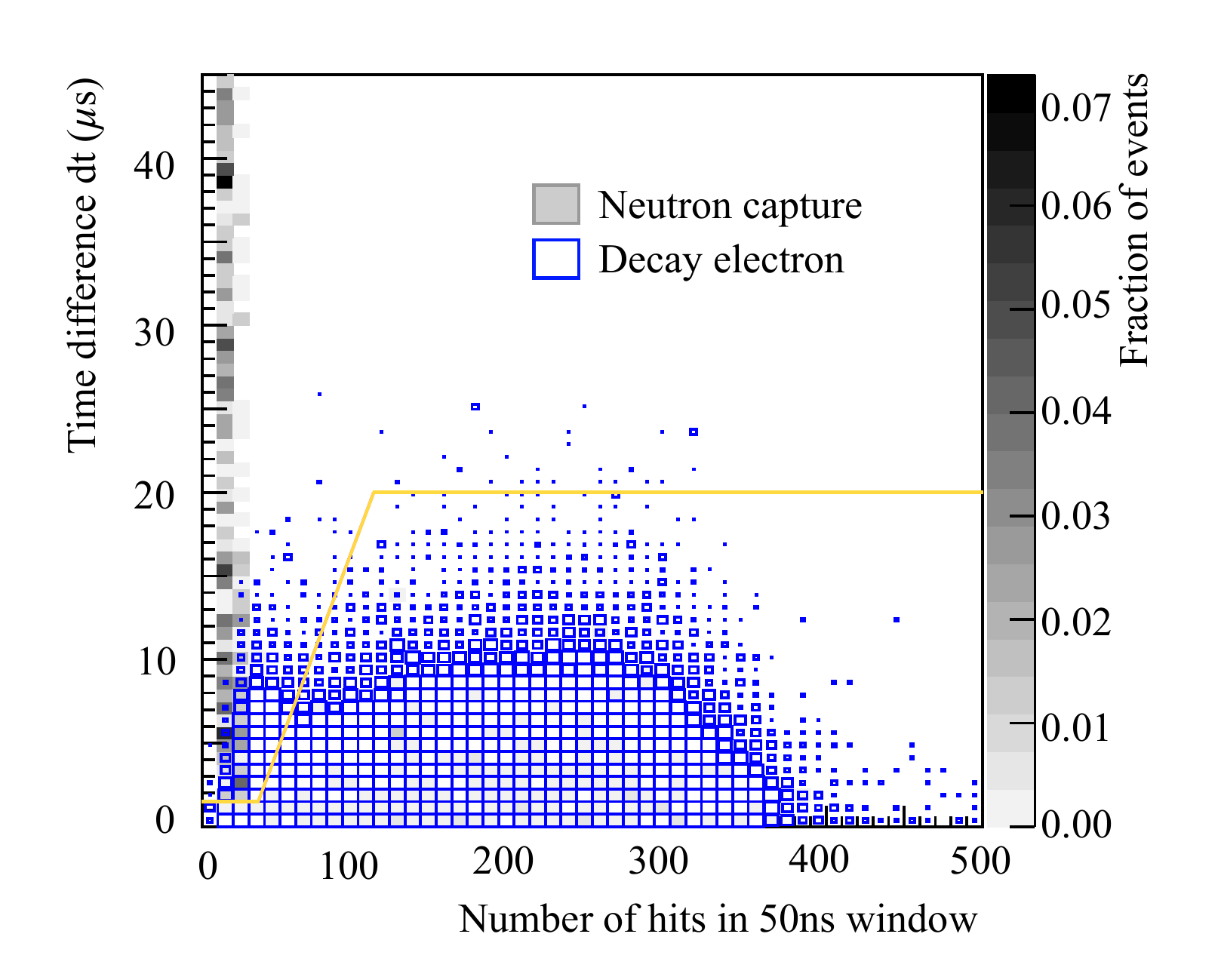}
        \caption{}
        \label{fig:decaye_combined_a}
    \end{subfigure}
    \hfill
    \begin{subfigure}{0.32\linewidth}
        \centering
        \includegraphics[width=\linewidth]{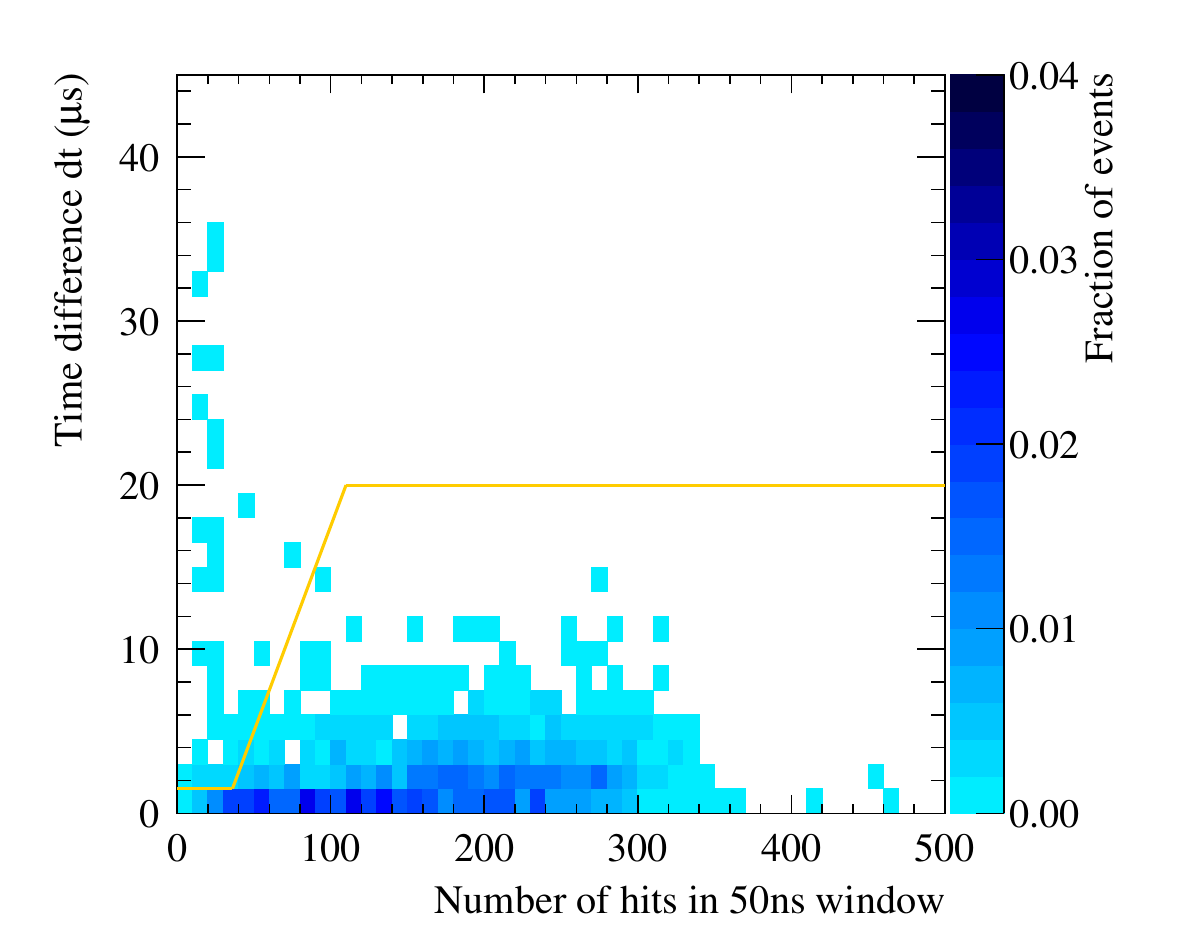}
        \caption{}
        \label{fig:decaye_combined_b}
    \end{subfigure}
    \hfill
    \begin{subfigure}{0.32\linewidth}
        \centering
        \includegraphics[width=\linewidth]{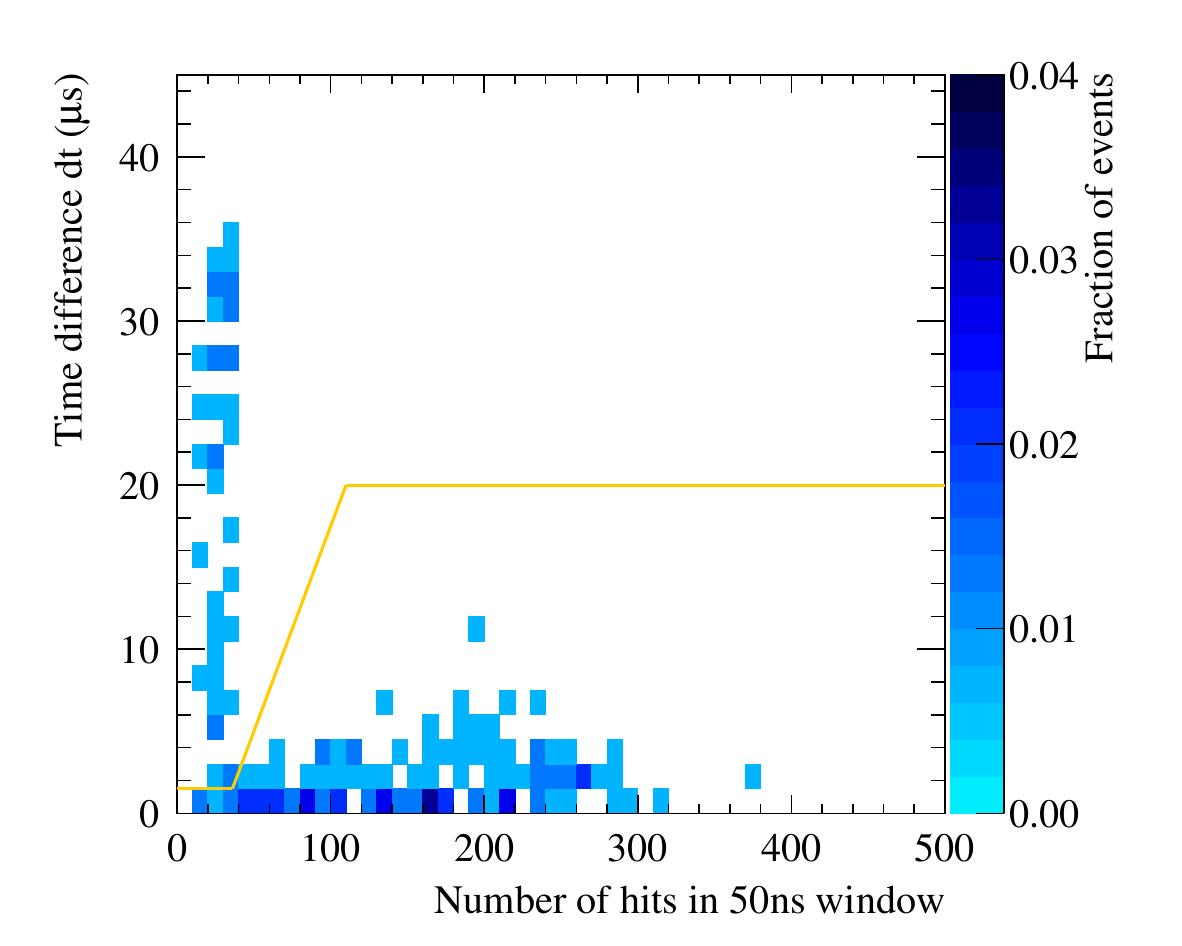}
        \caption{}
        \label{fig:decaye_combined_c}
    \end{subfigure}

   \caption{
The updated Michel electron selection cut (orange lines) shown in decay time dt ($\mu$s) versus the number of PMT hits within a 50 ns window around this time difference. 
(a) Monte Carlo comparison of neutron-capture events (gray) and Michel electron candidates (blue boxes), illustrating the discriminating power of the cut. 
(b) Performance of the cut on Pure-water phase FD data. 
(c) Performance on Gd-doped phase FD data. 
Events below the orange lines are classified as Michel electron candidates, while those above are preferentially rejected as neutron-capture events.
}
    \label{fig:decaye_combined}
\end{figure}

With 0.01\% Gd doping, the resulting increase in sensitivity to neutron-captures led to a 70\% increase in the misidentification rate of the existing Michel electron selection algorithm. To mitigate this effect, a new Michel electron selection cut was developed based on the time difference between the prompt neutrino event and the delayed PMT hit clusters, as a function of the number of delayed PMT hits within $50$~ns of the prompt neutrino time. As shown in Figure~\ref{fig:decaye_combined_a}, this phase space enables a strong discrimination between true neutron-capture events and Michel electron candidates. Neutron-captures are characterised by a relatively low number of PMT hits, approximately uniform across the lower decay time. The updated selection reduces the Michel electron candidate's misidentification rate to below $1\%$. Furthermore, it has a negligible impact on neutrino event selection from the pure-water phase (Figure~\ref{fig:decaye_combined_b}), ensuring consistency between the pure-water and the newly collected datasets. Resulting neutrino event rates in Gd-doped phase (Figure~\ref{fig:decaye_combined_c}) agree within 2--6\% from those obtained using the previous Michel electron selection algorithm.

\subsubsection{The new FD response model}
\label{sec:fd_sys}

Imperfect modelling of the detector's response introduces systematic uncertainties in the analysis. Since the number of beam events is low, T2K uses atmospheric neutrino events recorded at SK as control samples: any mismatch between data and simulation can then be attributed to detector mis-modelling. Assuming these mismatches can be reduced using bias and smearing corrections to the sample selection variables — such as PID and ring counting — a Bayesian Markov Chain Monte Carlo (MCMC) framework~\cite{MCMC-metro, MCMC-hasting}  is used to extract the parameter posterior distributions from the data and simulation. This procedure is referred to as the ``atmospheric fit''. Concretely, a selection variable $x$ is parametrised as
\begin{equation}
    x \rightarrow \alpha_{ij}\, x + \beta_{ij},
\end{equation}
where $\alpha_{ij}$ and $\beta_{ij}$ are the smearing and bias parameters respectively. The index $i$ runs over final-state topologies, including the dominant signal channels (e.g., single-ring $e/\mu$ events) as well as background topologies (e.g., $\pi^0/\pi^\pm$ events), while the index $j$ spans bins of visible energy.

In prior work~\cite{T2K:2021xwb, OA22shortPaper}, the results of the atmospheric fit were first encoded into an intermediate covariance matrix, from which values of shift and smearing parameters were randomly drawn and applied to the FD selection variables. The resulting fluctuation of event rates in each FD sample with respect to the nominal event rates were used to build the final covariance matrix, binned in reconstructed neutrino energy or lepton momentum — a procedure hereafter referred to as the ``toy MC procedure''. For this analysis, the intermediate matrix was removed: bias and smearing corrections were sampled directly from the posterior distributions of the atmospheric fit and applied to the FD simulation. This approach provides a more direct propagation of uncertainties and avoids the potential overestimation of errors introduced by the intermediate covariance-matrix step in earlier analyses.

These past analyses also used a relatively simple systematic model for the flux and cross-section nuissance parameters present in the atmospheric fit. Due to this, very large ad-hoc uncertainties ranging from 60--150\% were applied on background components of the FD samples, unconstrained in the atmospheric fit, during the toy MC procedure. In addition, no correlations between single- and multi-ring samples were considered. The bias and smearing parameters were also parametrised based on detector region instead of visible energy, as is done now.

In this analysis, event migrations between the single-ring and multi-ring samples are introduced, naturally inducing (anti-)correlations between the two categories. In addition, all ad hoc uncertainties~\cite{T2K:2021xwb} previously applied to nuisance parameters in the toy MC procedure have been removed, as they are now constrained directly by the improved flux and cross-section systematic model used in the atmospheric fit. 

\begin{figure}[htb!]
    \centering
    \includegraphics[width=0.8\linewidth]{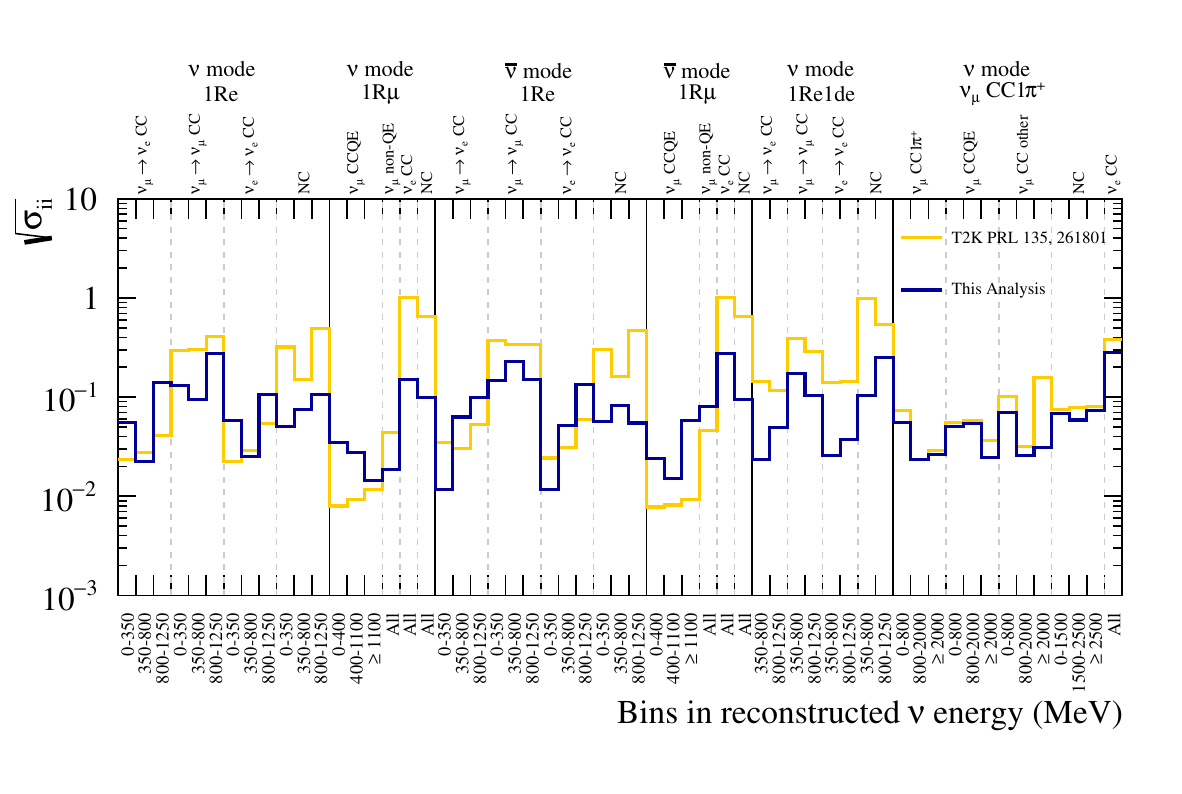}
    \caption{Fractional uncertainties for different FD samples, represented by the standard deviation of far detector error covariance matrix bins. The orange line shows the fractional uncertainties of the FD sample used in the previously published analysis~\cite{OA22shortPaper}, while the blue line shows the corresponding uncertainties used in this analysis.}
    \label{fig:skmatdiagonal}
\end{figure}

The impact of the updated FD mis-modelling uncertainties on the FD samples is shown as the standard deviation of each uncertainty bin in the covariance matrix in Fig.~\ref{fig:skmatdiagonal}, where blue and orange lines correspond to this analysis and the previous one respectively. The largest reduction in uncertainties is observed in the 1R$e$1d.e. sample, which is understood to arise from the removal of the intermediate covariance matrix step, which had previously led to an overestimation of uncertainties for this sample.

A further notable feature is the increase in uncertainties in some signal bins, which is attributed to the change in the shift and smearing parametrisation from detector-region–based to visible-energy–based modelling. This introduces additional correlations across reconstructed neutrino energy bins due to energy-dependent event migrations, leading to a redistribution of uncertainties across the signal phase space rather than a uniform reduction.

\begin{figure}[htb!]
\centering

\begin{subfigure}[t]{0.45\textwidth}
    \centering
    \includegraphics[width=\textwidth]{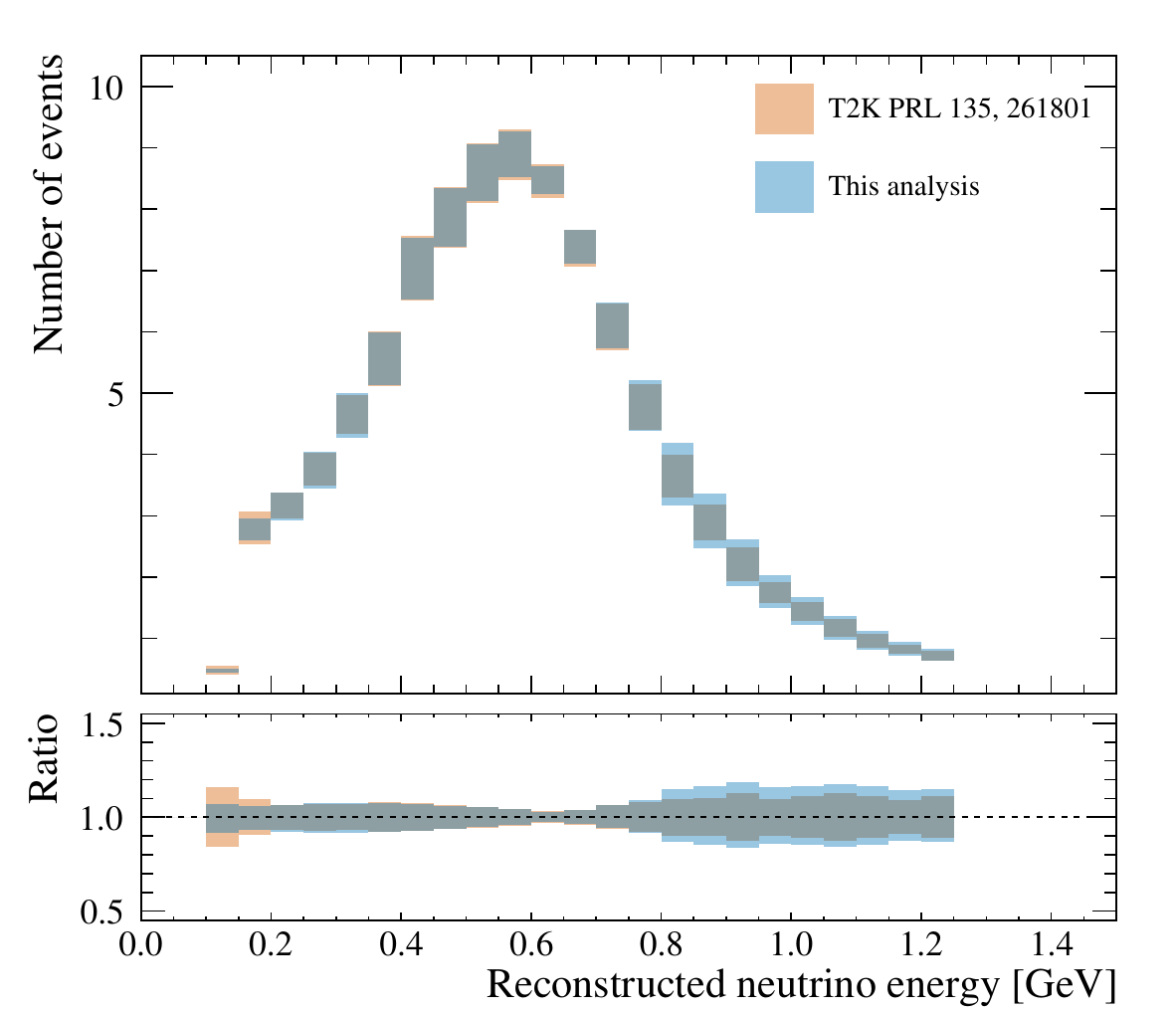}
    \caption{$\nu$ Mode 1R$e$ sample}
\end{subfigure}
\qquad
\begin{subfigure}[t]{0.45\textwidth}
    \centering
    \includegraphics[width=\textwidth]{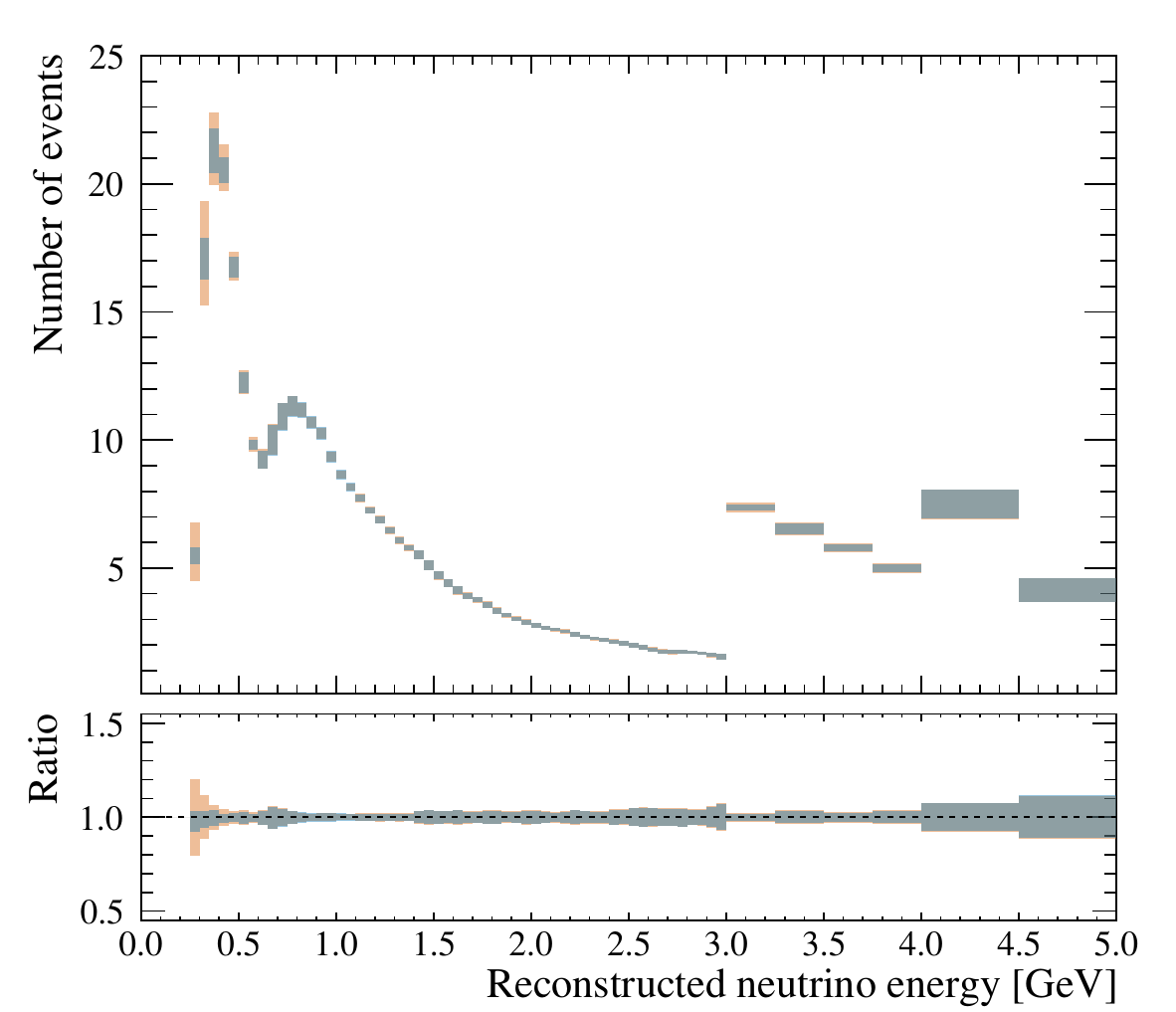}
    \caption{$\nu$ Mode 1R$\mu$ sample}
\end{subfigure}

\vspace{0.5em}

\begin{subfigure}[t]{0.45\textwidth}
    \centering
    \includegraphics[width=\textwidth]{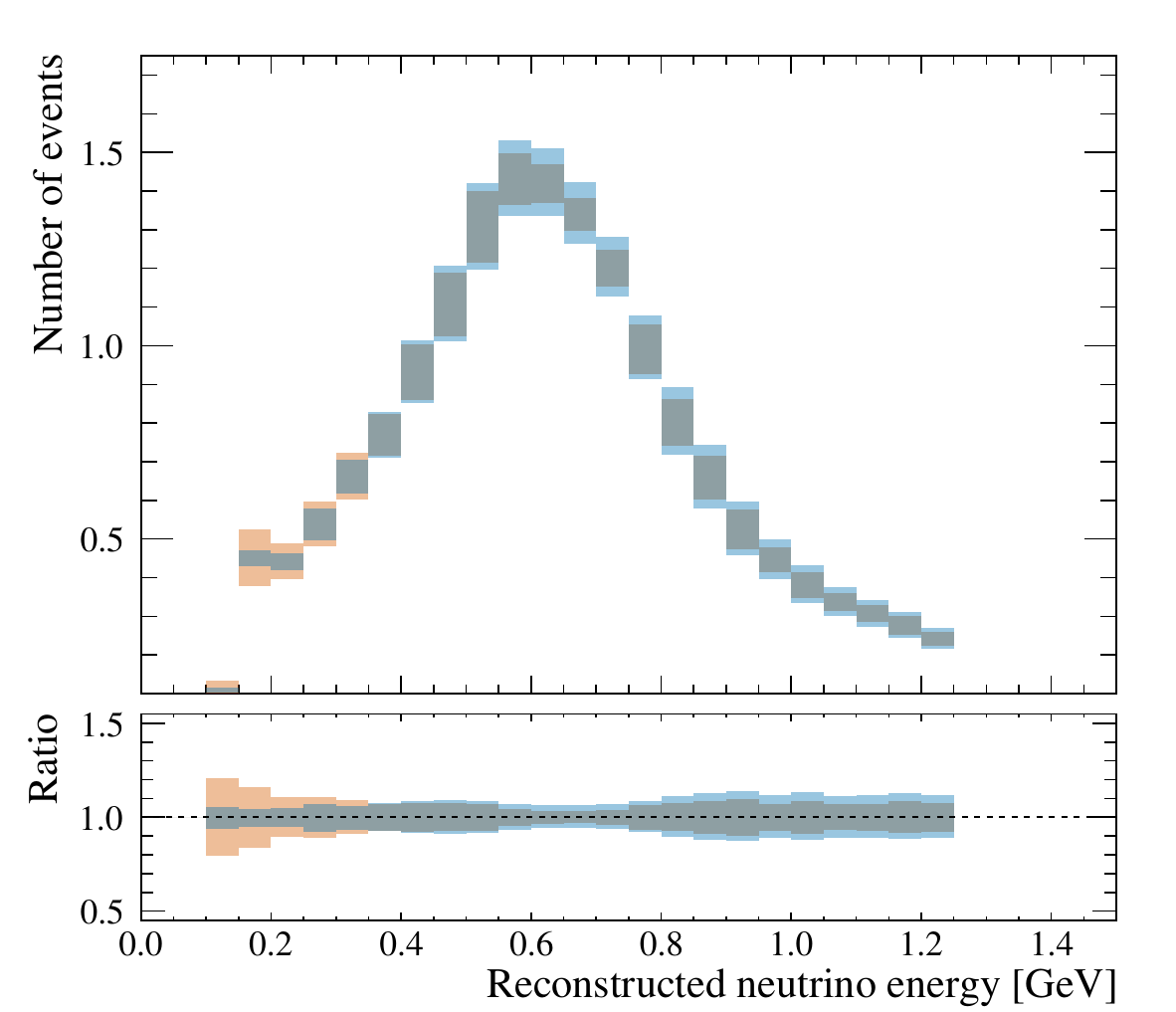}
    \caption{$\overline{\nu}$ Mode 1R$e$ sample}
\end{subfigure}
\qquad
\begin{subfigure}[t]{0.45\textwidth}
    \centering
    \includegraphics[width=\textwidth]{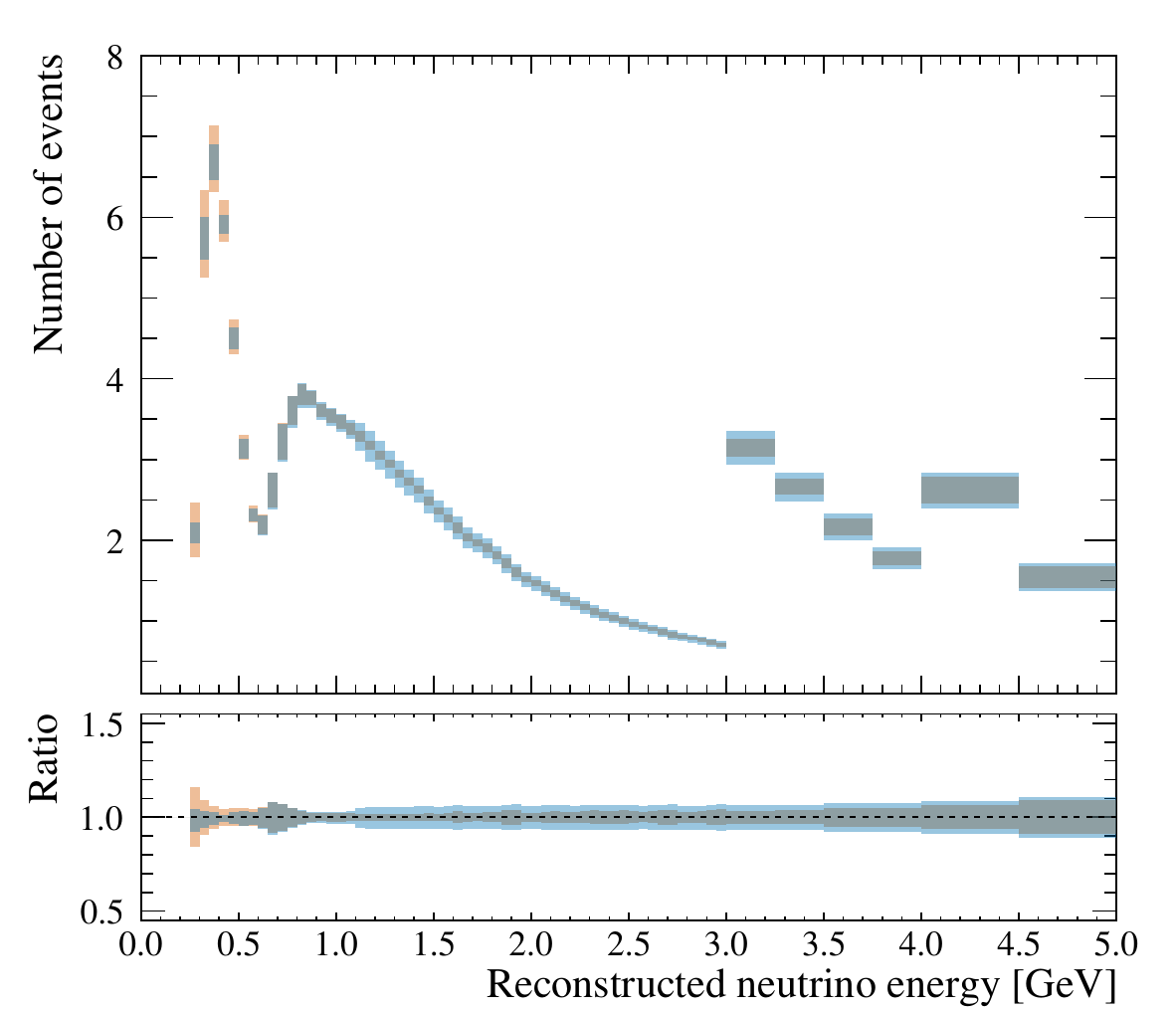}
    \caption{$\overline{\nu}$ Mode 1R$\mu$ sample}
\end{subfigure}

\vspace{0.5em}

\begin{subfigure}[t]{0.45\textwidth}
    \centering
    \includegraphics[width=\textwidth]{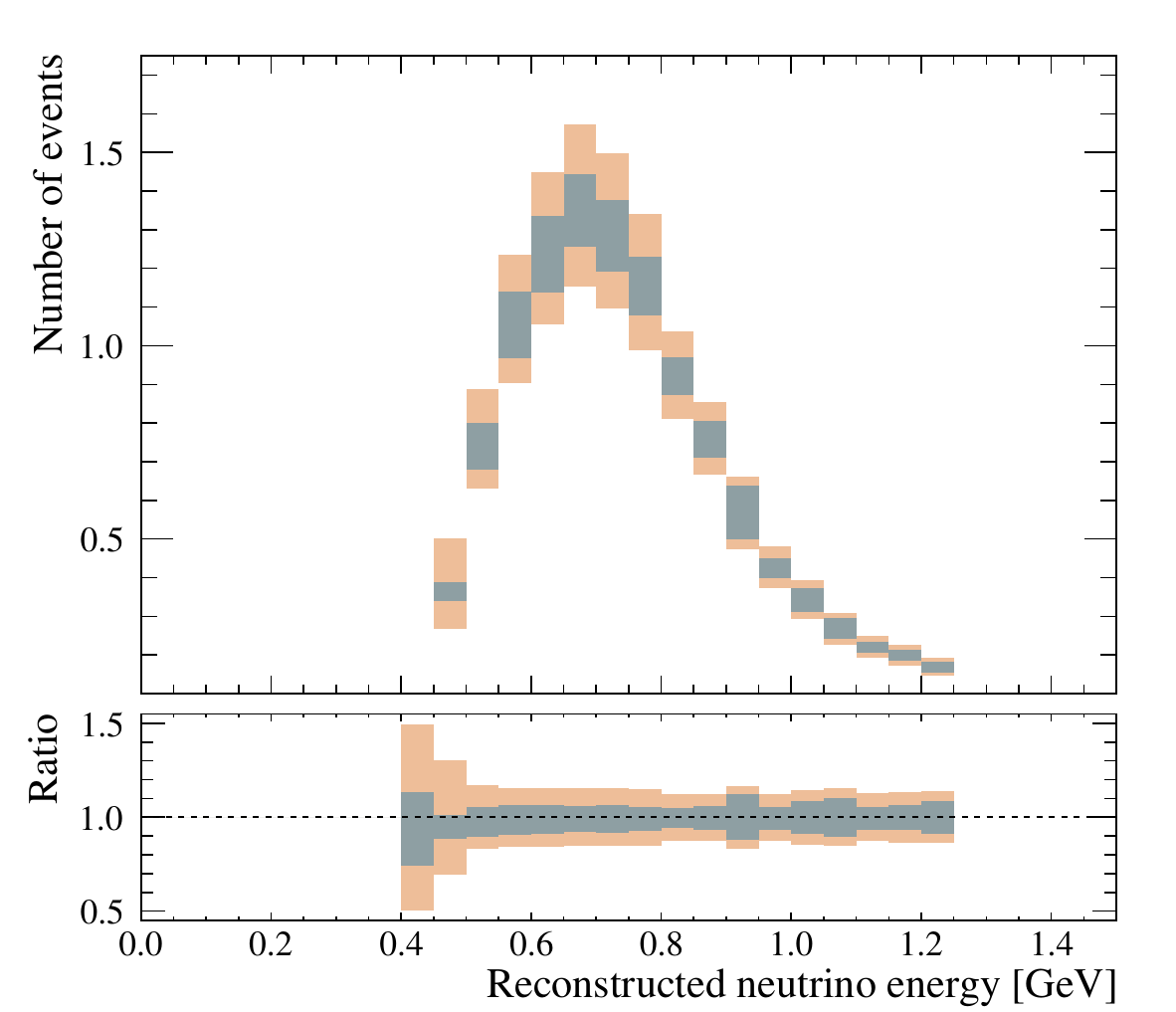}
    \caption{$\nu$ Mode 1R$e$1d.e sample}
\end{subfigure}
\qquad
\begin{subfigure}[t]{0.45\textwidth}
    \centering
    \includegraphics[width=\textwidth]{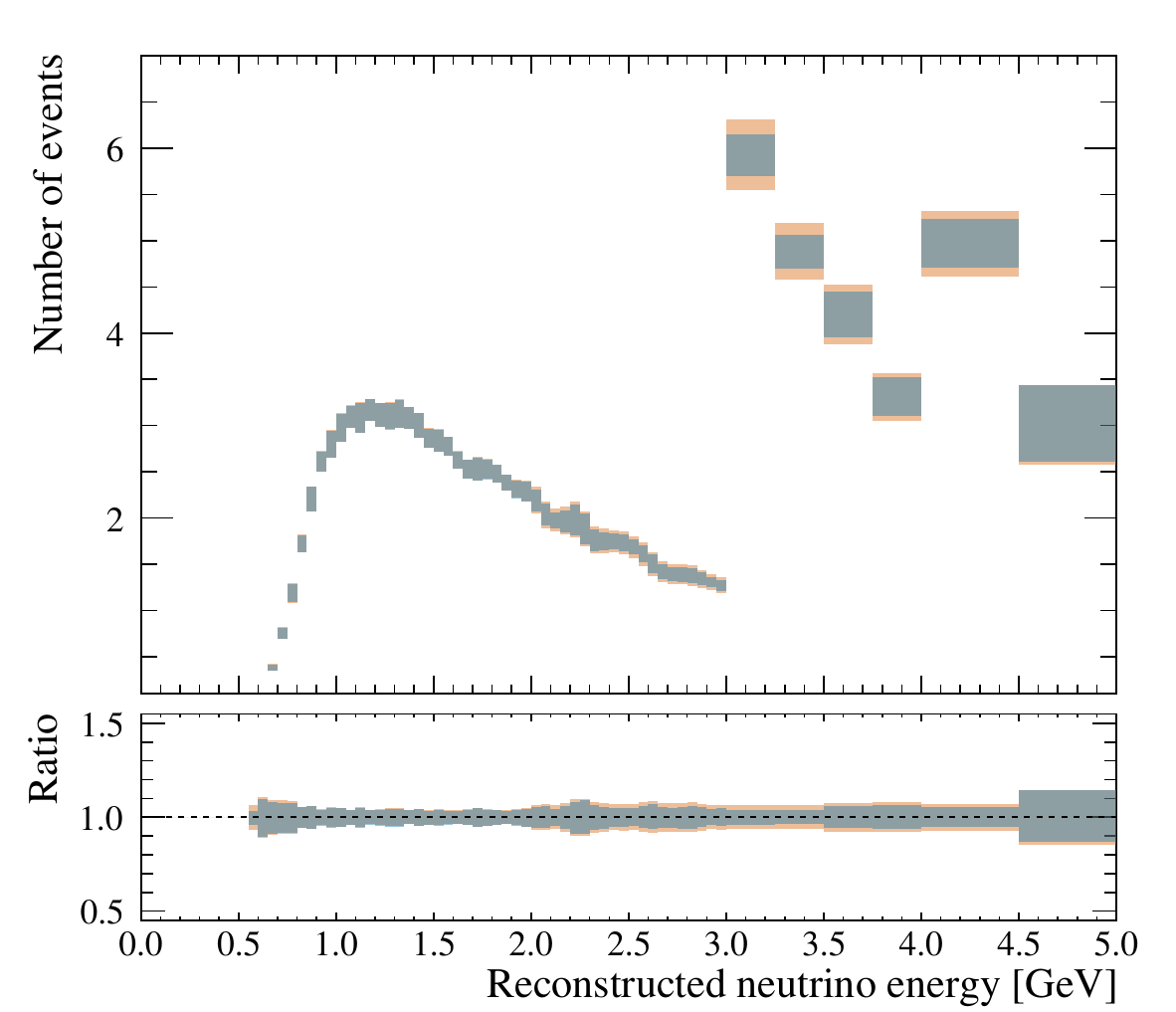}
    \caption{$\nu$ Mode $\nu_\mu$ CC$1\pi^+$-like sample}
\end{subfigure}

\caption{The change in uncertainties between the SK detector matrix used in the previous analysis (vermilion)~\cite{OA22shortPaper} and this analysis (blue).}
\label{fig:SKpred_newmat}
\end{figure}

The FD sample spectra error bands, obtained from the FD covariance matrix, are shown in Fig.~\ref{fig:SKpred_newmat}, comparing the impact of the previous matrix (vermilion) and the new matrix (blue). For the $\nu$- and $\overline{\nu}$-mode 1R$\mu$ samples, the reduction in background uncertainties leads to smaller prediction uncertainties at low reconstructed neutrino energies.

For the $\overline{\nu}$-mode 1R$e$ and 1R$\mu$ samples, uncertainties increase moderately at higher energies resulting from the increased uncertainties in some of the signal bins; however, the impact on the overall FD prediction remains small, and error reductions are still observed in the lowest energy bins.

\subsection{Systematic uncertainties}
The systematic uncertainties considered in this analysis are divided into three main sources: neutrino beam flux modeling, neutrino interaction modeling and the detector response. The detector response and some interaction model parameters, due to different materials at the ND and FD, are specific to either the ND or FD. Most flux and interaction parameters, however, are constrained by both detectors. Table \ref{tbl:percenterrors_simple} gives the breakdown of uncertainty on the number of events for each SK sample as found in the semi-frequentist fitter.

\begin{table}[htb!]
\caption{Uncertainty (in~\%) on the number of events in each SK sample distinguished by error source after the ND fit. Statistical uncertainties are shown for comparison. }		
	\label{tbl:percenterrors_simple}
\centering
\scalebox{0.9}{
\begin{tabular}{c || c c | c c || c c}
\toprule
&
\multicolumn{6}{c}{\textbf{Relative uncertainty (\%)}} \\
\cmidrule(lr){2-7}
&
\multicolumn{2}{c|}{$\nu$ Mode} &
\multicolumn{2}{c||}{$\bar{\nu}$ Mode} &
\multicolumn{2}{c}{$\nu$ Mode} \\
\makecell[c]{\textbf{Error source}} &
1R$e$ &
1R$\mu$ &
1R$e$ &
1R$\mu$ &
1R$e$1de &
$\nu_\mu$ CC1$\pi^+$-like \\
\midrule\midrule
Beam flux &
2.8 & 2.8 & 3.0 & 2.9 & 2.9 & 2.9 \\
Cross-section (ND constrained) &
3.8 & 3.6 & 3.5 & 3.5 & 4.3 & 3.0 \\
Flux \& Cross-section (ND constrained) &
2.9 & 2.8 & 2.7 & 2.6 & 3.7 & 2.2 \\
Cross-section (ND unconstrained) &
2.9 & 0.6 & 3.4 & 2.4 & 2.8 & 1.3 \\
SK detector &
2.7 & 1.4 & 5.1 & 3.6 & 4.3 & 2.9 \\
\midrule\midrule
\textbf{Total Syst.} &
4.9 & 3.2 & 6.7 & 5.0 & 6.3 & 3.9 \\
\hline
\textbf{Stat.} &
9.9 & 5.3 & 25.0 & 8.5 & 25.8 & 8.5 \\
\bottomrule
\end{tabular}
}
\end{table}

The ND constraint reduces the overall uncertainty from 12.1--17.4\% to 3.2--6.7\%. Compared to the previous analysis \cite{OA22shortPaper}, the flux and interaction modeling errors are similar, but, as explained in \ref{sec:fd_sys}, the SK far detector systematics have been updated and the error has been reduced in the 1R$e$1de sample from 13.3\% to 4.3\%. The overall systematic uncertainty is of a similar order of magnitude as the statistical error in the most populated samples (5.3\% statistical error on $\nu$ mode 1R$\mu$) and sub-dominant for the lower statistic samples (26\% statistical error on 1R$e$1de).

\subsection{Oscillation analysis}
\subsubsection{Fitting methods}
\label{sec:AnaMethod}
The Bayesian and semi-frequentist oscillation analysis methodologies are the same as in the previous analysis~\cite{OA22shortPaper}. They rely on a binned log-likelihood approach with binning dependent both on the selection and fitting methodology used. In the Bayesian analysis, all the $\mu$-like samples are binned purely in reconstructed neutrino energy and the $e$-like samples are binned in reconstructed neutrino energy and the angle of the outgoing lepton. In the semi-frequentist analysis, 1R$\nu_\mu$ samples in both $\nu$ and $\bar{\nu}$ modes are binned in reconstructed neutrino energy and outgoing lepton angle, while the $\nu_\mu$CC1$\pi^+$-like sample is binned in reconstructed neutrino energy only. The e-like samples are binned in the outgoing lepton momentum and angle.
All oscillation probabilities are calculated under PMNS 3-flavor hypothesis with a constant matter density of 2.6 g/cm$^{3}$. Fits are performed with flat priors on $\sin^{2}\theta_{23}$, $\Delta m^{2}_{32}$ and $\delta_{CP}$. Each neutrino mass ordering is given a prior probability of 50\%. $\Delta m^{2}_{21}$ and $\sin^{2}\theta_{12}$ are constrained with Gaussian priors which use the PDG~\cite{PDG2022} global best fit since these parameters are not well-constrained by T2K. The analyses are computed both without and with external constraints on $\sin^{2}\theta_{13}$, to respectively provide an independent measurement of $\sin^{2}\theta_{13}$ and maximize the constraints on \deltacp. For the results with external constraints, $\sin^{2}\theta_{13}$ is constrained with a Gaussian prior of $(2.20\pm0.07)\times 10^{-4}$ taken from~\cite{PDG2022}.%has been taken from reactor neutrino experiments \cite{Bak_2019, DoubleChooz:2014kuw, RENO:2018dro}{\color{red}(is it different from PDG24? Let's mention clearly it is taken from PDG2020. Check which PDG version. -> 2.20 is PDG22,  PDG24 reports 2.19.)}. 

\subsubsection{Impact of the new detector uncertainties and additional statistics}
\begin{figure}[h]
%\begin{center}
\includegraphics[width=7.5cm,keepaspectratio]{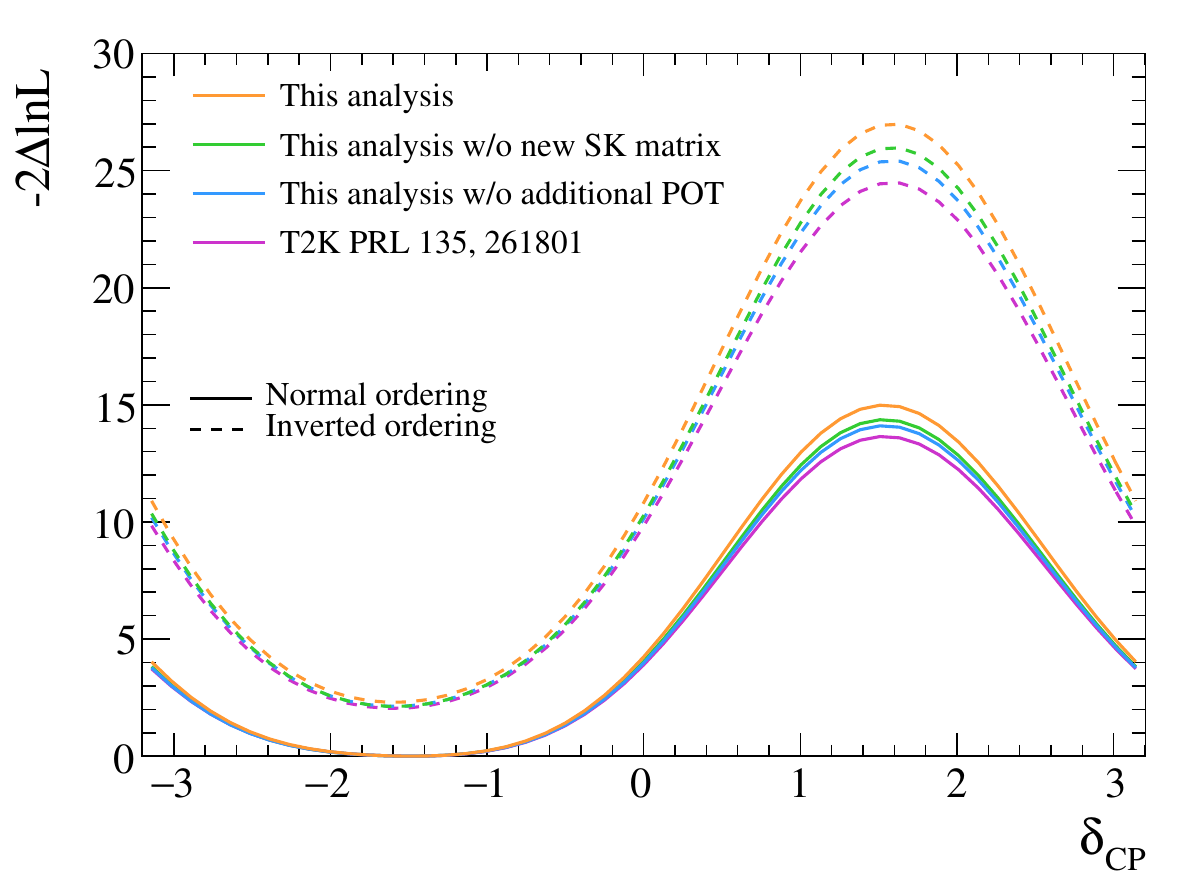}
\includegraphics[width=7.5cm,keepaspectratio]{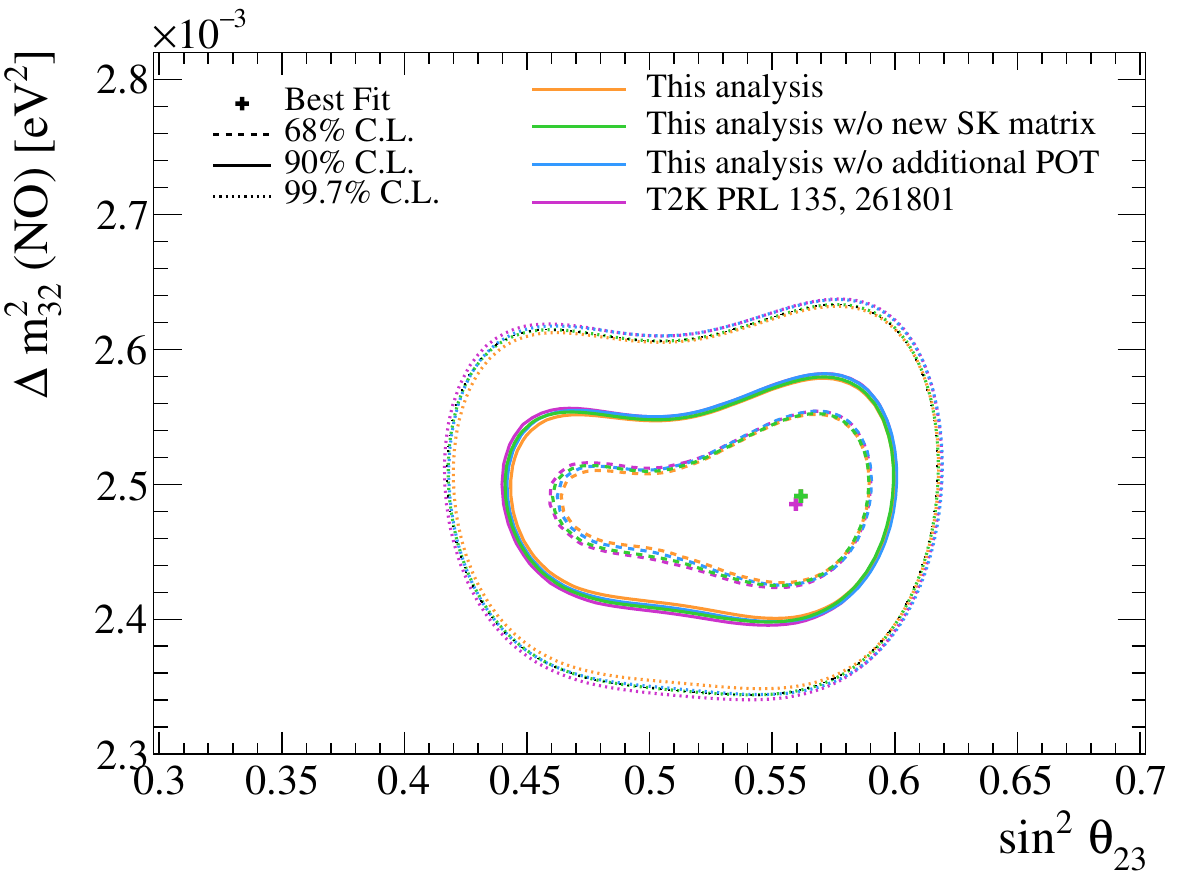}
%\end{center}
\caption{Sensitivity at A22 using the semi-frequentist fitter, with reactor constraints and with systematic parameters. The purple curve is the previous analysis sensitivity~\cite{OA22shortPaper}, the orange curve is this analysis sensitivity, the green curve is this analysis without the SK detector systematics improvement, and the blue curve is this analysis without the additional POT. The \deltacp sensitivity is given in normal (solid line) and inverted (dashed line) orderings. The $\Delta m^2_{32}$ (NO)/$\sin^2 \theta_{23}$ sensitivity is given for the normal ordering at various confidence levels.}
\label{fig:A22novelties}
%\end{center}
\end{figure}

The expected sensitivity of the analysis to oscillation parameters is evaluated for a single combination of true oscillation parameters described in Table~\ref{tab:AsimovA22}, that will be hereafter referred to as Asimov A22. It uses the last T2K semi-frequentist best~fits for the so-called atmospheric mass splitting squared which is $\Delta m^2_{32}$ in normal ordering and $\Delta m^2_{13}$ in inverted ordering, $\sin^2 \theta_{23}$ and \deltacp parameters. It also uses the values from~\cite{PDG2022} for $\Delta m^2_{12}$, $\sin^2 \theta_{12}$ and $\sin^2 \theta_{13}$.
Figure~\ref{fig:A22novelties} shows the sensitivity for the most impacted parameters: \deltacp (left) in both mass orderings and in the $\Delta m^2_{32}$ (NO)/$\sin^2 \theta_{23}$ plane (right) in normal ordering. The sensitivity of this analysis is also evaluated with the former analysis~\cite{OA22shortPaper} statistics and detector uncertainties, in order to assess the impact of each update separately. It shows that the sensitivity has been improved compared to the previous analysis, and that the impact of the additional POT is comparable to the impact of the reduced errors in the SK detector matrix.

\begin{table}[htb]
\caption{Set of oscillation parameter values used for sensitivity studies.}
\centering
\begin{tabular}{ c | c  }
\hline
Parameter set                           & Asimov A22                     \\
\hline                                                                       
$\deltacp$                       & -1.601       \\       
$\sin^2 \theta_{13}$                & 0.0220                     \\
$\sin^2 \theta_{23}$                & 0.561                    \\
$\sin^2 \theta_{12}$                & 0.307                    \\
$\Delta m^2_{32}~(\text{eV}^2)$ & $2.494\times10^{-3}$ \\
$\Delta m^2_{21}~(\text{eV}^2)$ & $7.53\times10^{-5}$  \\
\hline\hline
\end{tabular}
\label{tab:AsimovA22}
\end{table}

\subsection{Simulated data studies}
\label{sec:FDS}
As described in \cite{OA22shortPaper}, a set of 19 simulated data studies (SDS) is performed to test the robustness of the analysis, and especially, the near-to-far detector extrapolation, against alternative nuclear and cross-section models as well as other effects. The semi-frequentist fitter and analysis are used to quantify these studies. For each simulated data set, the biases and uncertainties on the $\Delta m^2_{32}$ (NO)/$\vert\Delta m^2_{31}\vert$ (IO) and $\sin^{2}\theta_{23}$ parameters are computed, and compared to the Asimov systematic and total uncertainties. When these biases or uncertainty changes overcome our criteria, the above uncertainties on parameters are modified. More explicitly, no action is required when the bias on the center of the $2\sigma$ interval is less than 50\% and the change in the $2\sigma$ interval width is less than 10\%.
%and the impact on the physics conclusion on CP violation is checked. The criteria 
It is found that an additional error on $\Delta m^2_{32}$ (NO)/$\vert\Delta m^2_{31}\vert$ (IO) is necessary to be applied in the form of a Gaussian smearing with width $S = 3.1\times10^{-5}$~eV$^2$. This value is comparable to the total value of the error on the parameter, emphasizing the importance of these studies. No action is needed for 
reporting the $\sin^{2}\theta_{23}$ result, as none of the studies exceed the criteria we have established.\\
Finally, the impact on the physics conclusion on CP violation of each data set is tested and presented in Section~\ref{sec:Results}. 

\section{Results and discussion}
\label{sec:Results}

\subsection{Event rate comparisons}
Table~\ref{tab:run1-11_data} shows the number of selected events at FD compared to simulation for various values of \deltacp. The data p-values have also been evaluated fixing the model oscillation parameters to their best fit values, while varying the other parameters of the model within their prior uncertainties provided by the near detector fit. One observes a general agreement between our model and data, and a mild preference of our e-like samples for a large CP violation. %\textcolor{green}{It's not the right place where to discuss CP violation}.
Moreover, a deficit in the data 1R$\mu$ channel is observed compared to our model corresponding to a rate p-value of 0.02, which is less significant than 0.05 after applying Bonferroni correction~\cite{Bayer:2020pva}. Moreover, the shape in reconstructed energy of the events in this sample shows a relative agreement with our model.
%\textcolor{green}{Don't we need the systematic errors of table 2 to compute the p-values? (Mauro)} 

\begin{table}
    \caption{Number of expected simulated events at various $\deltacp$ values for each SK selected sample for all runs. The number of data events is shown for comparison. The rate and rate+shape p-values are also shown.}
     \footnotesize
    \centering
    \begin{tabular}{|c|c|c|c|c|c|c|c|c|c|}
   
\hline
Mode & Selection & $\delta_{CP}$ & $\delta_{CP}$ & $\delta_{CP}$ & $\delta_{CP}$ & $\delta_{CP}$ & Data & rate & rate+shape \\
 & & $=-\pi/2$ & $=0$ & $=\pi/2$ & $=\pi$ & $=-2.08$ & & p-value & p-value \\
% & & & & & & & & \\
\hline
\hline
$\nu$ & 1R$\mu$ & 417.2 & 416.3 & 417.1 & 418.2 & 419.5 & 357 & 0.02 & 0.33 \\
&$\nu_{\mu}$CC1$\pi^+$-like & 123.9
& 123.3 & 123.9 & 124.4 & 123.3 & 140 & 0.25 & 0.43\\
& 1R$e$ & 113.2 & 95.5 & 78.3 & 96.0 & 112.1 & 102 & 0.48 & 0.12 \\
& 1R$e$1de & 10.0 & 8.8 & 7.2 & 8.4 & 9.9 & 15 & 0.73 & 0.11\\
%  & 140 \\
\hline
$\bar\nu$ & 1R$\mu$ & 146.7 & 146.3 & 146.7 & 147.1 & 147.0 & 137 & 0.57 & 0.82\\
& 1R$e$ & 17.6 & 20.0 & 22.2 & 19.7 & 18.0 & 16 & 0.47 & 0.64\\

\hline
\end{tabular}

    \label{tab:run1-11_data}
\end{table}
\subsection{Oscillation parameters results}
\begin{figure}
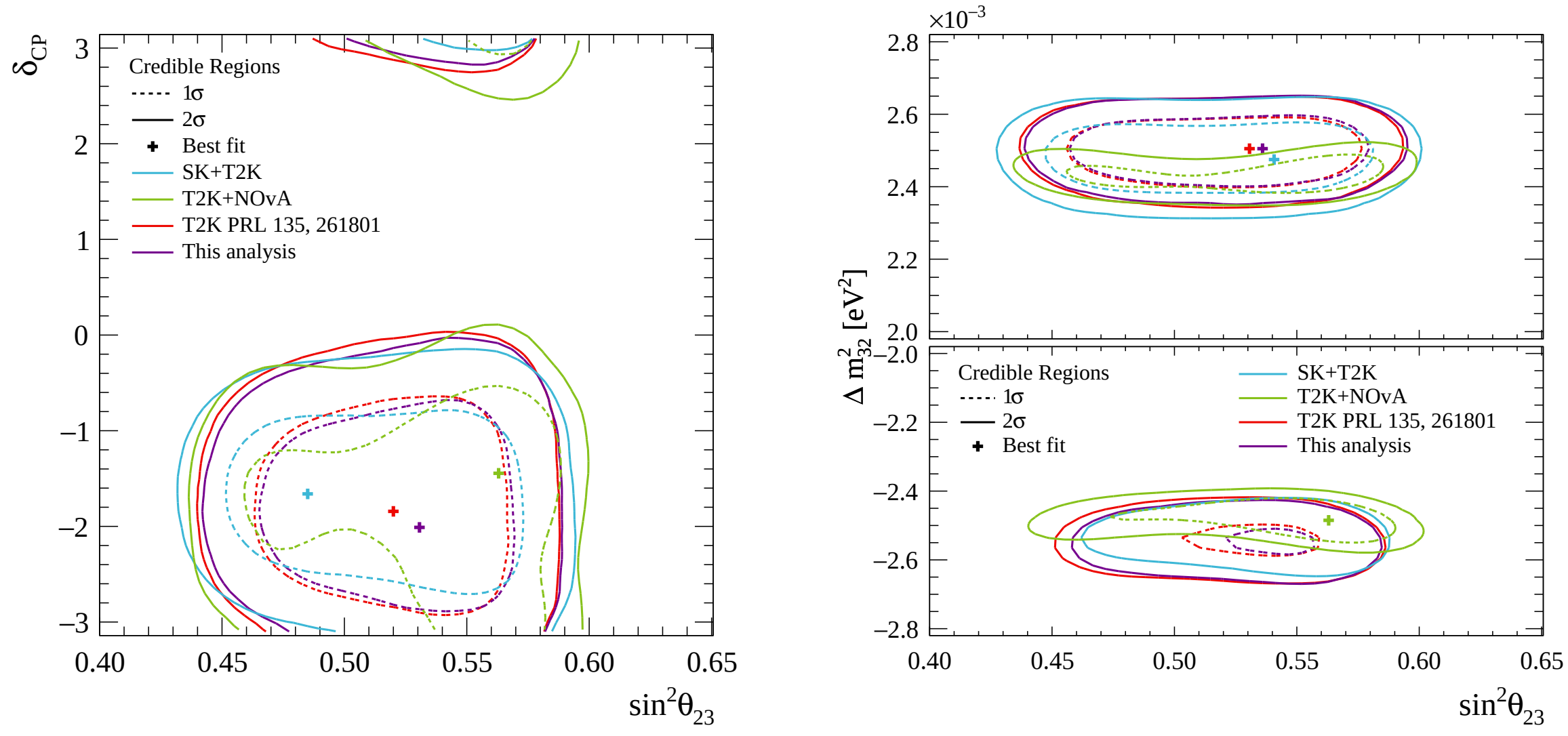

    \centering
    \begin{tabular}{cc}    
    \includegraphics[width=0.5\linewidth, page=2]{OA_Figures/UpdatedPaperPlots/ContourComparisons_2D_MOMarg_wRC.pdf} &
    \includegraphics[width=0.5\linewidth, page=3]{OA_Figures/UpdatedPaperPlots/ContourComparisons_2D_MOMarg_wRC.pdf}
    
    \end{tabular}
    %\caption{ Comparison of Bayesian credible intervals produced by the Bayesian fitter between this analysis (purplish) and the previous T2K analysis (reddish) \cite{OA22shortPaper} with best fit points marked with a +. The appearance channel contours (left) show a slight tightening of both the 1$\sigma$ and 2$\sigma$ credible intervals. The best fit value (+ marker) is roughly stationary for both analyses. The disappearance channel (right) results show some small shifts in best fit contours, although the best fit points remain consistent. In both the plots, contours and best fits from SK+T2K and T2K+NOvA joint analysis {\color{red}(need citation)} are also overlaid.
    %bumpiness due to a lack of smoothing when producing the results.}
    \caption{ Comparison of credible intervals produced by the Bayesian fitter between this analysis (purple) and the previous T2K analysis (red) \cite{OA22shortPaper} with best fit points marked with a +. The ``appearance channel'' contours (left) show a slight tightening of both the 1$\sigma$ and 2$\sigma$ credible intervals. The best fit value is roughly stationary for both analyses. It should be noted that the credible intervals are marginalized over both mass ordering for this figure only. The ``disappearance channel'' (right) results show some small shifts in best fit contours, although the best fit points remain consistent. In both plots, contours and best fits from SK+T2K~\cite{JointFit} and T2K+NOvA~\cite{T2KNOvA2025} joint analyses are also overlaid.}
    \label{fig:oa22_23_contour_comp}
\end{figure}

\begin{figure}[h]
%\begin{center}
\includegraphics[width=7.5cm,keepaspectratio]{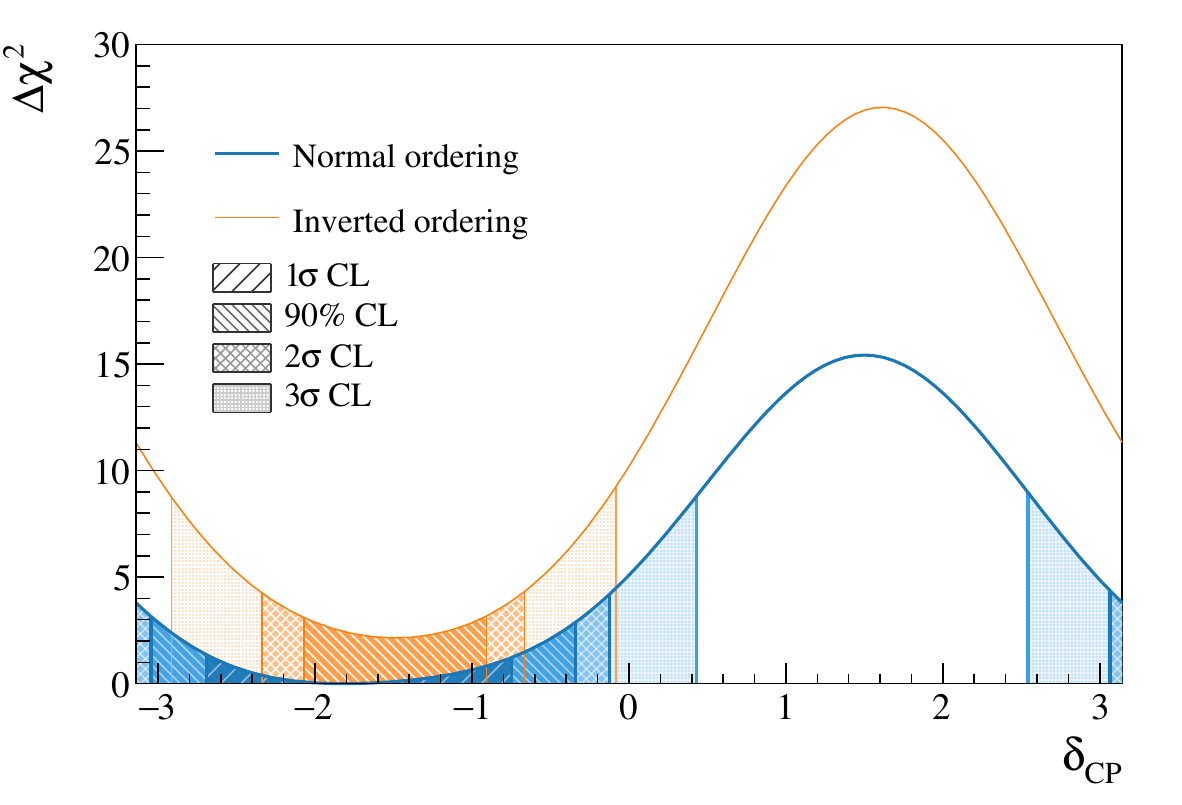}
\includegraphics[width=7.5cm,keepaspectratio]{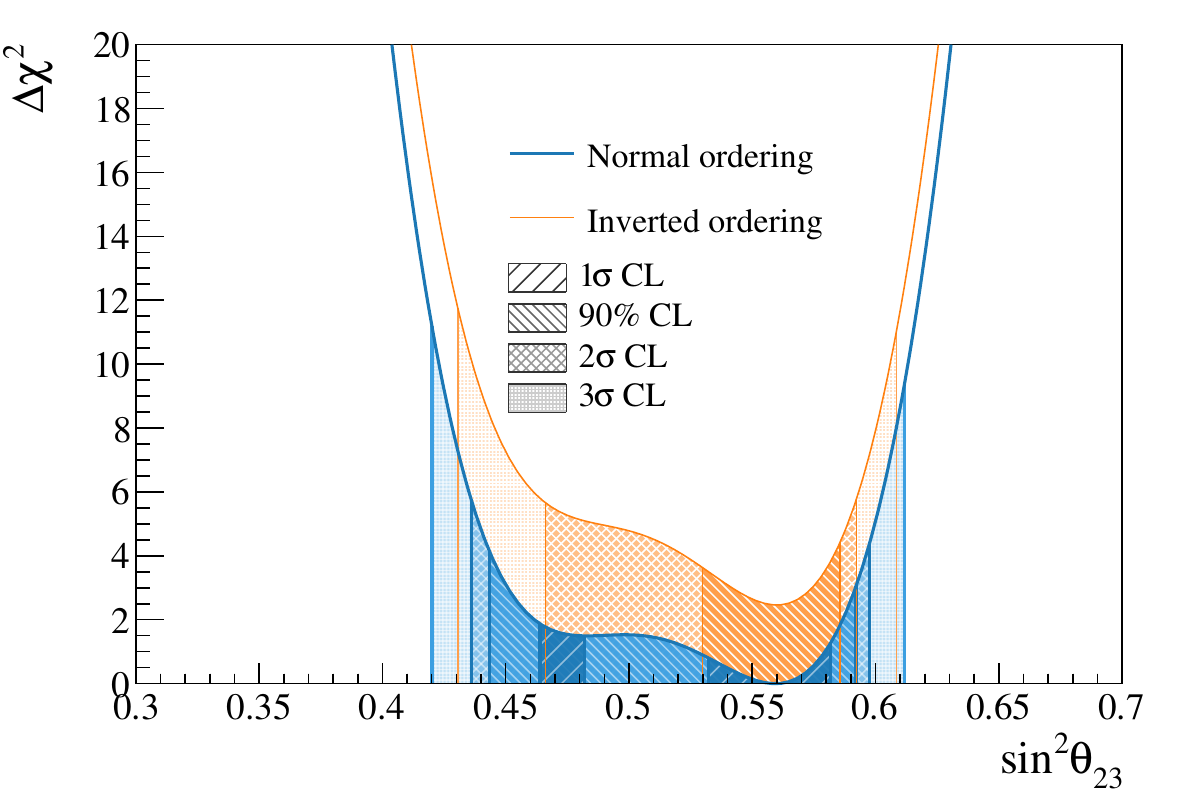}
%\end{center}
\caption{Feldman-Cousins confidence intervals reported on the 1D likelihood curve from the semi-frequentist data fit, using the reactor constraints. Results for \deltacp are shown on the left plot and for $\sin^{2}\theta_{23}$ on the right plot. Both normal (blue) and inverted (orange) ordering scenarios are shown. }
\label{fig:PthetaFC}
\end{figure}

The results of applying both Bayesian and semi-frequentist methods are first shown using the reactor constraint on $\theta_{13}$. The result of the Bayesian analysis is shown in Figure~\ref{fig:oa22_23_contour_comp} which details the credible interval contours in the $\sin^{2}\theta_{23}$–$\deltacp$ plane. Note that these credible intervals are produced through marginalization over both neutrino mass-ordering \textit{i.e.} both signs of $\Delta m^{2}_{32}$. It allows simpler comparison across several experiments, while T2K has sub-leading sensitivity to mass ordering compared to $\deltacp$. Figure~\ref{fig:oa22_23_contour_comp} also presents the credible intervals in the $\sin^{2}\theta_{23}$–$\Delta m^{2}_{32}$ plane under both mass ordering hypotheses separately. The semi-frequentist analysis results, shown in Figure~\ref{fig:PthetaFC}, rely on the Feldman-Cousins~\cite{FC} method to build the confidence intervals for \deltacp and $\sin^{2}\theta_{23}$. The best fit points and $1-3\sigma$ intervals are reported in Table~\ref{tab:best_fit_summary_MOSplit} for the Bayesian and Table~\ref{tbl:PthetaRes} for the semi-frequentist results for both neutrino mass ordering hypotheses. To place this analysis in a broader context, contours from the previous T2K analysis, as well as two recent joint analyses—T2K with SK \cite{JointFit} and T2K with NOvA \cite{T2KNOvA2025}—are also overlaid in Figure~\ref{fig:oa22_23_contour_comp}. It should be noted, however, that the T2K dataset used in both joint analyses corresponds to the dataset described in \cite{T2K:2023smv}.\\
The CP conservation hypothesis is excluded by the frequentist confidence intervals at the $90\%$ level and lies outside the Bayesian corresponding $90\%$ credible intervals. The preferred value of \deltacp lies close to the maximal CP-violating value of $-\pi/2$ radians. As shown in Figure~\ref{fig:oa22_23_contour_comp}, both these results are consistent and represent a refinement over the previous analysis~\cite{OA22shortPaper}, driven by both the increased POT and upgrades to the FD systematic uncertainty model. As presented in Section~\ref{sec:FDS}, this exclusion robustness is also tested using 19 simulated data sets. Out of the 19, two fail to exclude CP conservation at $\deltacp = \pm\pi$ at a 90\% confidence level by a very small margin. The first set consists of modifying the single pion production at low $Q^{2}$ based on external measurements from MINER$\nu$A~\cite{MinervaSPPQ2}. The second has been driven by a $<2\sigma$ excess of data in the lepton low momentum region ($< 200$~MeV/c) of the 1R$e$1de sample~\footnote{This set has been constructed by doubling the number of events having a momentum $< 200$~MeV/c in 1R$e$1de sample for the simulation, allowing to retrieve an agreement with data in this momentum region}. Moreover, both analysis results are compatible with maximal mixing within $2\sigma$ confidence level, while slightly favoring the upper octant for $\sin^{2}\theta_{23}$, which is also in agreement with our previous analysis.\\
The $\deltacp$ contours are also in excellent agreement with the joint analyses—T2K with SK and NOvA, especially when the mass-ordering is marginalized out. Though the joint fit with SK (resp. NOvA) slightly prefers lower (resp. higher) octant, our new constraints on $\sin^{2}\theta_{23}$ also show agreement with respect to both joint fits. Finally, our $\Delta m^2_{32}$ (NO)/$\vert\Delta m^2_{31}\vert$ (IO) parameter is measured with a precision close to 2\%, and shows excellent agreement with the T2K+SK joint fit, preferring a slightly higher mass squared difference than the joint fit with NOvA. Note that T2K is also sensitive to $\sin^{2}\theta_{13}$, and its measurement without the reactor constraint is shown in Table~\ref{tbl:t13}.

\begin{table}[h]
\centering
\caption{Central values and credible intervals for oscillation parameters from Bayesian analysis. Central values are taken from the 2D disappearance ($\sin^{2}\theta_{23}$--$\Delta m^{2}_{32}$) and appearance ($\sin^{2}\theta_{13}$--$\deltacp$) posterior distributions and credible intervals (C.I.) from the 1D posterior distributions. Results are given for T2K + reactor experiment constrained fit.}
\renewcommand{\arraystretch}{0.75}
\resizebox{\textwidth}{!}{
\begin{tabular}{c|c|c|c|c|c|c}
\hline
Parameter & $\deltacp$ [rad.] NO & $\deltacp$ [rad.] IO & $\sin^{2}\theta_{23}$ NO & $\sin^{2}\theta_{23}$ IO & $\Delta m^2_{32}$ NO & $\Delta m^2_{32}$ IO\\
\hline \hline
Central value & -2.18 & -1.34 & 0.536 & 0.541 & 2.505 & -2.535\\
\hline
1$\sigma$ C.I. & [-2.78,-1.2] & [-1.94,-0.79] & [0.485,0.560] & [0.495,0.565] & [2.445,2.555] & [-2.60,-2.490]\\
90\% C.I. & [-3.09,-0.63] & [-2.36,-0.47] &[0.465, 0.575] & [0.47, 0.575] & [2.415,2.590] & [-2.64,-2.46]\\
2$\sigma$ C.I. & [ -$\pi$, -0.31] $\cup$ [3.04, $\pi$] & [-2.57,-0.31] &[0.455,0.58] & [0.46,0.58] & [2.395,2.610] &  [-2.655,-2.440] \\
3$\sigma$ C.I. & [ -$\pi$, 0.42] $\cup$ [2.46, $\pi$] & [-3.09,0.16] & [0.435,0.6]& [0.44,0.6] & [2.345,2.665] &  [-2.710,-2.390]\\
\hline
\end{tabular}}
\label{tab:best_fit_summary_MOSplit}
\end{table}

% trim={0 15pt 0 0},
% trim={0 0 0 15pt},

\begin{table}[h]
\centering
\caption{Semi-frequentist confidence intervals, using Feldman-Cousins for  $\sin^{2}\theta_{23}$ and \deltacp, in both normal (NO) and inverted (IO) mass orderings. The intervals for other parameters use the constant $\Delta \chi^{2}$ method. Results are given for T2K + reactor experiment constrained fit.}
%and{\arraystretch}{0.75}
\resizebox{\textwidth}{!}{\begin{tabular}{c|c|c|c|c|c|c}
\hline
Parameter & $\deltacp$ [rad.] NO & $\deltacp$ [rad.] IO & $\sin^{2}\theta_{23}$ NO & $\sin^{2}\theta_{23}$ IO & $\Delta m^2_{32}$ NO & $\Delta m^2_{13}$ IO \\
\hline\hline
Best fit & -2.08 & -1.41 & 0.568 & 0.567 & 2.521 & 2.486 \\
\hline
%1$\sigma$ C.L. & [-2.69,-0.75] &  & [0.464,0.482] $\cup$ [0.532,0.582]  &  & [2.471, 2.560] & [2.443, 2.531] \\
%90\% C.L. & [-3.04,-0.34] & [-2.07,-0.91] &[0.443,0.592] & [0.530,0.586]  & [2.443, 2.588] &	[2.414, 2.559] \\
%2$\sigma$ C.L. & [ -$\pi$, -0.13] $\cup$ [3.06, $\pi$] & [-2.34,-0.67] &[0.436,0.598] & [0.466,0.592] & [2.427, 2.604] & 	[2.399, 2.575] \\
%3$\sigma$ C.L. & [ -$\pi$, 0.43] $\cup$ [2.54, $\pi$] & [-2.92,-0.08] & [0.420,0.612]& [0.431,0.609] & [2.382, 2.649] & 	[2.355, 2.619]  \\
1$\sigma$ C.L. & [-2.69,-0.75] &  & [0.464,0.482] $\cup$ [0.532,0.582]  &  & [2.441, 2.590] & [2.403, 2.561] \\
90\% C.L. & [-3.04,-0.34] & [-2.07,-0.91] &[0.443,0.592] & [0.530,0.586]  & [2.413, 2.618] &	[2.384, 2.589] \\
2$\sigma$ C.L. & [ -$\pi$, -0.13] $\cup$ [3.06, $\pi$] & [-2.34,-0.67] &[0.436,0.598] & [0.466,0.592] & [2.397, 2.634] & 	[2.369, 2.605] \\
3$\sigma$ C.L. & [ -$\pi$, 0.43] $\cup$ [2.54, $\pi$] & [-2.92,-0.08] & [0.420,0.612]& [0.431,0.609] & [2.352, 2.679] & 	[2.325, 2.649]  \\
\hline
\hline
\end{tabular}}
\label{tbl:PthetaRes}
\end{table}

\begin{table}[h]
\centering
\caption{$\sin^{2}\theta_{13}$ central values and intervals for T2K-only fit under normal and inverted ordering hypothesis. Both Bayesian credible intervals and semi-frequentist confidence intervals using constant $\Delta \chi^{2}$ are shown.}
%\resizebox{0.9\linewidth}{!}{
\begin{tabular}{c|cc|cc}
\hline
Fitter    & \multicolumn{2}{c|}{Bayesian} & \multicolumn{2}{c}{Semi-Frequentist} \\ \hline\hline
Parameter & \multicolumn{1}{c|}{$\sin^{2}\theta_{13}$ NO}   &  $\sin^{2}\theta_{13}$ IO  & \multicolumn{1}{c|}{$\sin^{2}\theta_{13}$ NO}       &    $\sin^{2}\theta_{13}$ IO    \\ \hline\hline
Best fit  & \multicolumn{1}{c|}{0.0234}    &  0.0256  & \multicolumn{1}{c|}{0.0278}       &    0.0310    \\ \hline
1$\sigma$ & \multicolumn{1}{c|}{[0.0203, 0.0287]}    & [0.0221, 0.0315]   & \multicolumn{1}{c|}{[0.0209, 0.0296]}       &   [0.0236, 0.0328]     \\
90\%      & \multicolumn{1}{c|}{[0.0175, 0.0324]}    & [0.0203, 0.0352]   & \multicolumn{1}{c|}{[0.0188, 0.0334]}       &   [0.0213, 0.0366]     \\
2$\sigma$ & \multicolumn{1}{c|}{[0.0165, 0.0343]}    & [0.0193, 0.0371]   & \multicolumn{1}{c|}{[0.0177, 0.0358]}       &   [0.0201, 0.0388]     \\
3$\sigma$ & \multicolumn{1}{c|}{[0.0165, 0.0343]}    & [0.0156, 0.0436]   & \multicolumn{1}{c|}{[0.0150, 0.0434]}       &   [0.0171, 0.0460]     \\ \hline\hline
\end{tabular}%
%}
\label{tbl:t13}
\end{table}

\subsection{Mass ordering constraint}

\begin{figure}[ht]       
\centering
\includegraphics[width=0.8\textwidth ]{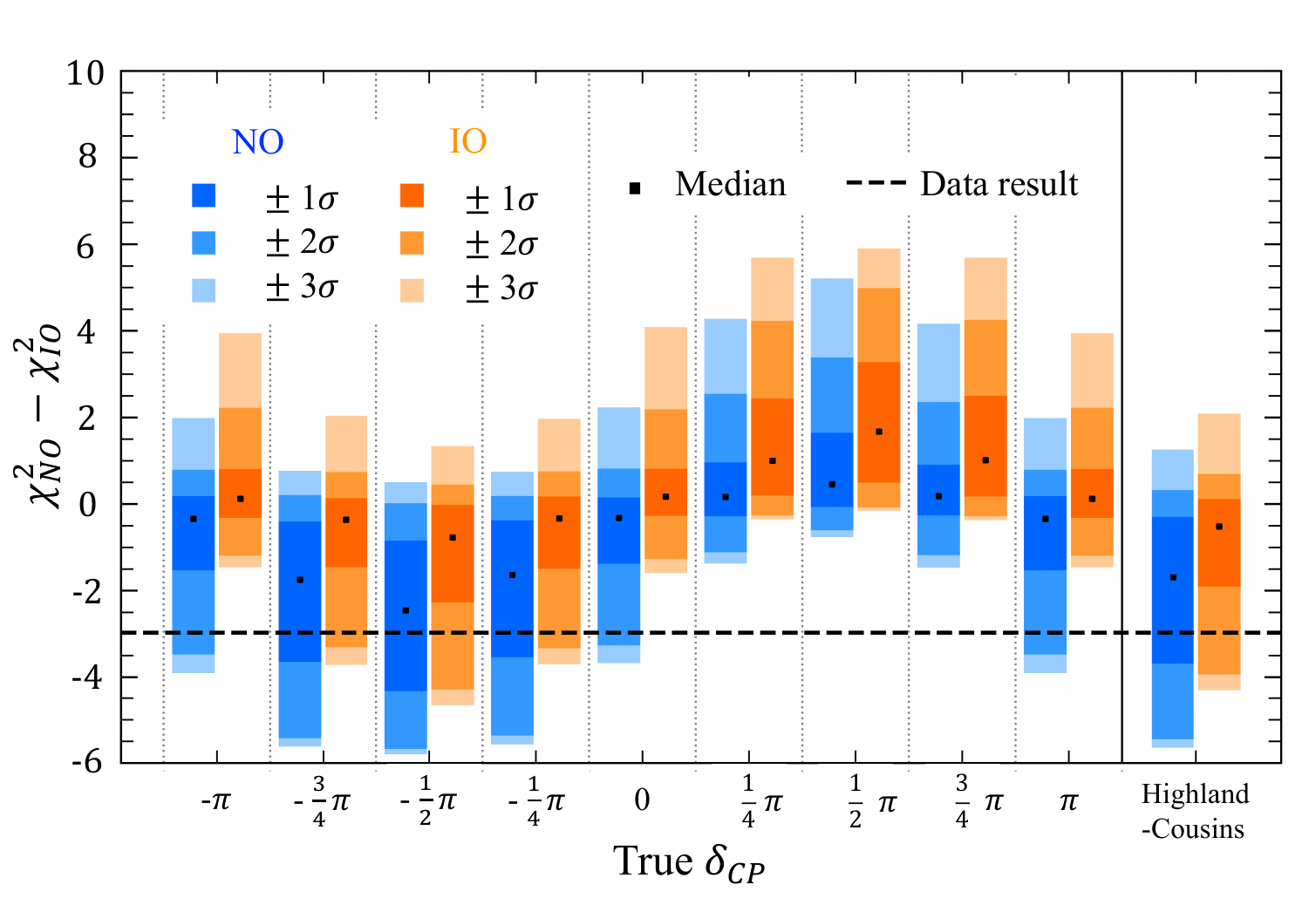}
\caption{Ability to distinguish between the mass orderings for two different toy generation methods: fixed and non-fixed $\delta_{CP}$ (Highland-Cousins method). The bar indicates the central range of values of $\Delta \chi^2$ containing 68\%, 95\% and 99.7\% of the values obtained for the pseudo-experiments generated at each case. The dashed line corresponds to the value obtained in the data fit ($\Delta \chi^2=-2.972$).}
\label{fig:CLS_main}
\end{figure}

The T2K data-fit result favors the normal mass ordering (NO). To quantify the significance of this preference in the semi-frequentist fit, we calculated the p-values ($p_{0}$) and confidence levels~\cite{CL0,CL1,JUNK1999435} (CLs) for the inverted mass ordering (IO) hypothesis. The test statistic employed is \(\Delta \chi^2 = \chi^2 (\text{NO}) - \chi^2 (\text{IO})\). In this study, ten sets of 50~000 pseudo-experiments were generated, where systematic parameters were sampled from prior distributions, while oscillation parameters were drawn from the data-fit posterior: nine sets correspond to fixed \(\deltacp\) values, and one set allows \(\deltacp\) to be sampled from the data fit posterior through the posterior Highland-Cousins method~\cite{HighlandCOUSINS}. 
The results are summarized in Figure \ref{fig:CLS_main}. The first nine columns represent different assumed true values of \(\deltacp\), ranging from \(-\pi\) to \(\pi\), to illustrate the impact of the degeneracy between the CP-violating phase and the mass ordering. The final column presents the results obtained using the Posterior Highland-Cousins method, where all oscillation parameters, including \(\deltacp\), were sampled from the posterior distribution of the data fit.
In the case of non-fixed \(\deltacp\), the p-value and confidence for inverted ordering hypothesis are $p_{0}(\text{IO}) = 0.0603$ $(1.88\sigma)$ and $\text{CLs(IO)} = 0.213$ $(1.25\sigma)$.

In addition to the semi-frequentist studies, the Bayesian fit with reactor constraint also produced a Bayes factor of $\mathcal{B}(NO/IO)=3.35$. It highlights that there is insufficient evidence in our result to make physics conclusions about mass ordering.

\FloatBarrier
\section{Conclusions}
The T2K collaboration has presented the measurement of the PMNS matrix \ssqthtwothree, $\Delta m^2_{32} $, \ssqthonethree and \deltacp parameters using $21.4(16.3) \times 10^{20}$ POT in (anti-)neutrino mode at the SK far detector. Compared to previous analysis~\cite{OA22shortPaper}, this analysis uses an additional neutrino data set of $1.7 \times 10^{20}$ POT collected with SK filled with Gadolinium (Gd). Furthermore, significant improvements were added to the SK detector selection and systematic error treatment, resulting in a reduction of our detector systematic uncertainties.\\
\\
This data analysis is compatible with \ssqthtwothree maximal mixing, with a slight preference for the upper octant, and the normal ordering of neutrino masses. These results also confirm the exclusion of CP-conservation with a $90\%$ confidence level. This exclusion robustness is tested against 19 simulated data sets, among which two fail to exclude CP conservation at $\deltacp = \pm\pi$ at a 90\% confidence level by a very small margin. Finally, the parameter $1\sigma$ confidence levels are reported under normal ordering hypothesis and using the external constraint from reactor experiments: $\Delta m^{2}_{32} = 2.521^{+0.037}_{-0.050} \times 10^{-3}$~eV$^{2}$ using constant $\Delta \chi^{2}$ intervals; $\sin^{2} \theta_{23} = 0.568^{+0.014}_{-0.036}$ (lower octant range 0.464--0.482 at $1\sigma$) and $\deltacp = -2.08^{+1.33}_{-0.61}$ using Feldman–Cousins corrected intervals. The $1\sigma$ confidence levels are also reported for $\sin^{2} \theta_{13} = 2.78^{+0.18}_{-0.69} \times 10^{-2}$ without the reactor constraint, using constant $\Delta \chi^{2}$ method.
%\textcolor{red}{Nuance with FDS results ?}

%$\delta_{CP} = 0$ is excluded at the 2$\sigma$ confidence level, while $\delta_{CP}=\pi$ at 1$\sigma$. This is consistent with previous T2K analyses. 
%Similar with last analysis~\cite{OA2022}, a Jarlskog invariant value of zero is excluded at more than $2\sigma$ credible level assuming a flat prior in $\delta_{CP}$, and just below $2\sigma$ with a flat prior in $\sin \delta_{CP}$.
%In the upcoming analyses...
% I think in the next analysis, the main imporvements will be 4pi selection at ND, new SK sample: nue multi-ring (merged with the current nue1de), FD binning optimization and improvements to the interaction uncertainty model. And then for the future analyses : ND280 upgrade and new data runs
\section*{Acknowledgment}

The T2K collaboration would like to thank the J-PARC staff for superb accelerator performance. We thank the CERN NA61/SHINE Collaboration for providing valuable particle production data. We acknowledge the support of MEXT,   JSPS KAKENHI (JP16H06288, JP18K03682, JP18H03701, JP18H05537, JP19J01119, JP19J22440, JP19J22258, JP20H00162, JP20H00149, JP2O0J20304) and bilateral programs (JPJSBP120204806, JPJSBP120209601),  Japan; NSERC, the NRC, and CFI, Canada; the CEA and CNRS/IN2P3, France; the Deutsche Forschungsgemeinschaft (DFG, German Research Foundation) 397763730, 517206441, Germany; the NKFIH (NKFIH 137812 and TKP2021-NKTA-64), Hungary; the INFN, Italy; the Ministry of Science and Higher Education (2023/WK/04) and the National Science Centre (UMO-2018/30/E/ST2/00441, UMO-2022/46/E/ST2/00336 and UMO-2021/43/D/ST2/01504), Poland;  the RSF (RSF 24-12-00271) and the Ministry of Science and Higher Education, Russia; MICINN  (SEV-2016-0588, PID2022-136297NB-I00 /AEI/10.13039/501100011033/ FEDER, UE, PID2021-124050NB-C31, PID2020-114687GB-I00,  PID2019-104676GB-C33), Government of Andalucia (FQM160, SOMM17/6105/UGR) and the University of Tokyo ICRR's Inter-University Research Program FY2023 Ref. J1, and ERDF funds and CERCA program, and University of Seville grant Ref. VIIPPIT-2023-V.4, and Secretariat for Universities and Research of the Ministry of Business and Knowledge of the Government of Catalonia and the European Social Fund (2022FI\_B 00336), Spain; the SNSF and SERI (200021\_185012, 200020\_188533, 20FL21\_186178I), Switzerland; the STFC and UKRI, UK; the DOE, USA; and NAFOSTED (103.99-2023.144,IZVSZ2.203433), Vietnam. We also thank CERN for the UA1/NOMAD magnet, DESY for the HERA-B magnet mover system, the BC DRI Group, Prairie DRI Group, ACENET, SciNet, and CalculQuebec consortia in the Digital Research Alliance of Canada, GridPP and the Emerald High Performance Computing facility in the United Kingdom, the WestGrid, SciNet and CalculQuebec consortia in Compute Canada, and the CNRS/IN2P3 Computing Center in France and NERSC (HEP-ERCAP0028625). In addition, the participation of individual researchers and institutions has been further supported by funds from the ERC (FP7), “la Caixa” Foundation  (ID 100010434, fellowship code LCF/BQ/IN17/11620050), the European Union’s Horizon 2020 Research and Innovation Programme under the Marie Sklodowska-Curie grant agreement numbers 713673 and 754496, and H2020 grant numbers  RISE-GA822070-JENNIFER2 2020 and RISE-GA872549-SK2HK; the JSPS, Japan; the Royal Society, UK; French ANR grant number ANR-19-CE31-0001, ANR-21-CE31-0008, ANR-24-CE31-2754, ANR-24-CE93-0006-01 and ; and  Sorbonne Université Emergences programmes; the SNF Eccellenza grant number PCEFP2\_203261;  the VAST-JSPS (No. QTJP01.02/20-22);  and the DOE Early Career programme, USA. For the purposes of open access, the authors have applied a Creative Commons Attribution license to any Author Accepted Manuscript version arising.


%\bibliographystyle{ptephy}
%\bibliography{sample}
%
% once the .bbl file has been generated then place the text in your article.

\vspace{0.2cm}
\noindent

%This is added by T. Yoneya (editor-in-chief) on 2020/07/09.

\let\doi\relax

%without this code before the command "\begin{thebibliography}{}" , an error will be %flagged. When the bibliography is provided as separate .bib file, then this code %should be placed above the commands "\bibliographystyle{}" and "\bibliography{}" %inside the main TeX file. 

\bibliography{bibliography}

\end{document}